\documentstyle{l-aa}
\input psfig
\begin{document}
   \thesaurus{12         
              (11.03.1;  
               11.17.3;  
               12.03.3;  
               12.04.1;  
               12.07.1)} 

   \title{Identification of a high redshift cluster in the field
of Q2345+007 through deep BRIJK' photometry\thanks{Based
on observations collected at the
Canada-France-Hawaii Telescope at Mauna Kea, Hawaii, USA}}

   \author{R. Pell\'o\inst{1}
\and J. M. Miralles\inst{1}
\and J.-F. Le Borgne\inst{1}
\and J.-P. Picat\inst{2}
\and G. Soucail\inst{1}
\and G. Bruzual\inst{3}
          }

   \offprints{R. Pell\'o, roser@obs-mip.fr}

   \institute{Observatoire Midi-Pyr\'en\'ees, Laboratoire
d'Astrophysique de
    Toulouse, URA 285, 14 Avenue E. Belin, F-31400 Toulouse, France
\and
Observatoire Midi-Pyr\'en\'ees, URA 1281, 14 Avenue E. Belin, 
F-31400 Toulouse, France
\and
    Centro de Investigaciones de Astronom\'\i a, AP 264, 5101-A M\'erida,
Venezuela
             }

\date{Received November 9, 1995, Accepted , 2001}

\maketitle 
\markboth{R. Pell\'o et al.: Identification of a high-z cluster in
Q2345+007 }{}



\begin{abstract}
This paper describes new results on the identification of the 
complex gravitational lens responsible for the double
quasar Q2345+007. A gravitational shear field was
detected recently about 45\arcsec\ away from the QSO, centered on
an excess of faint blue galaxies. The redshift distribution
is still unknown, so the mass and the photometric properties of the 
deflector are still a matter of debate.
We present deep photometric data obtained in the
near-IR (J and K'), which are
used together with preexisting optical BRI photometry to build
spectral energy distributions
for all the objects of the field, and to derive a photometric
redshift estimate by comparison with synthetic spectrophotometric 
data. We propose a
statistical method to analyse the redshift distribution, based
on the cumulative histogram of the redshift ranges allowed for the
different objects. An excess of galaxies at a redshift of $z \simeq
0.75$ is clearly detected in the field of Q2345+007, with a 2D 
distribution showing a maximum
located at the center of the weak shear field. Besides, the redshift
inferred for this cluster is also compatible with that found for an
absorption system in the spectrum of the B component of the quasar.
We interpret this overdensity of objects
as a distant cluster of galaxies responsible for the gravitational
shear field. Two other redshift concentrations are studied:
$z = 0.28$ which corresponds to the spectroscopic redshift of
three galaxies but for which no strong excess of objects is identified, 
and $z \simeq 1.2$, where
an excess of galaxies is also detected, but with a rather 
smooth 2D distribution over our field of view. We also discuss
the existence of other possible excesses of galaxies at 
redshift planes compatible with the absorption systems detected 
in the spectra of the QSOs. Most cluster-member candidates at 
$z \simeq 0.75$ are undergoing a star-formation process or 
are burst systems where the star formation stopped between 1
and 3 Gyr ago.

\end{abstract}

\section{Introduction}
Since its discovery (Weedman et al. 1982), the nature of the
double quasar Q2345+007 has been a controversial matter. The spectra of
both quasars and their redshifts are very similar ($z = 2.15$),
therefore the gravitational lensing
hypothesis was suggested at once (Foltz et al. 1984). The distance
between the two quasars is $7.1\arcsec$, a value so high to be 
explained by a single galaxy deflector that
Subramanian \& Chitre (1984) suggested a double lens model. 
The best model they found was that of a galaxy in the
central part of a cluster at a redshift close to 1. Nevertheless, the
first deep visible images of the field did not show any trace of a lens
candidate around the two images of the QSO 
(Tyson et al. 1986). This negative result leaded Tyson et al. to
conclude that the mass to light ratio for the lens might be at least
1000, a value which is compatible with the hypothesis of a massive halo
of dark matter suggested by Narayan et al. (1984). Conversely,
Steidel \& Sargent (1990) and Weir \& Djorgovski (1991)
concluded that the double quasar is probably a physical pair.
Nevertheless, after a careful analysis of both spectra, Steidel \&
Sargent (1991) showed that they are really similar in many respects
(the redshift, the emission lines as well as the absorption lines) and
they definitely favour the gravitational lens hypothesis. Besides, the
metal absorption systems identified in both spectra, at 
redshifts ranging from $z = 0.754$ to $z =1.98$, 
allow to suspect the presence of lens candidates at these
redshifts. The close neighbourhood of the quasar was also studied in
the near IR, up to a magnitude $K' = 20$, in a recent paper by McLeod et
al. (1994), but the detection of a lens candidate was still negative.
   
All the works reviewed before to search for the lens deflector assume 
implicitely that the main deflector must be located close to the two QSO 
images. In a different approach, Bonnet et al. (1993) reported the 
detection of a shear field due to the weak gravitational 
distortion of the background sources by a mass distribution compatible
with that of a cluster of galaxies. Its center is located about $45\arcsec$ away 
from the double QSO, close to two bright galaxies identified at a 
redshift of $z=0.28$. Moreover, several arclets candidates were also 
identified, reinforcing the hypothesis of a strong deflecting mass.
Deeper optical imaging confirmed the suspected 
excess of faint blue  galaxies associated with the lens (Mellier et al. 1994, 
Fischer et al. 1994). All these authors argued that its redshift could be higher than 
1 because of the low surface brightness and color indices, and even as 
large as 1.5 if the excess is associated with the absorbing systems of the 
QSOs. But a precise value for the redshift cannot be inferred from their
observations.
A secondary small clump of blue galaxies was found close
to the double quasar. As there are absorption lines in
the spectra of both quasars at $z = 1.49$, this value for the
redshift was considered as the most convincing one. 

The galaxies in excess in the field of Q2345+007 are so faint that 
they are hardly observable 
in spectroscopic mode, and the redshift estimate of the deflecting 
agents is still a matter of debate. In this paper, we propose a different
approach by using deep near-IR J and K' photometry 
of the field in complement of the existing deep B,R and I photometry.
We show that from this spectral information, which covers a wide wavelength 
range, and a careful analysis of the photometric errors, 
it is possible to infer a more constrained 
{\it photometric redshift} for each object in the field. The observed
spectral energy distribution (hereafter SED) is compared with the 
predictions for different redshifts and spectromorphological 
types of galaxies. If detectable, a cluster along the line-of-sight must
appear as an excess of objects at a particular value of the redshift 
distribution with respect to an empty field. A discussion about the properties
and limits of such a photometric-redshift technique can be found in 
previous papers by Couch et al. (1983) and Ellis et al. (1985). 
Recently, Belloni et al. (1995) have applied successfully a similar photometric
method to study the population of galaxies in a distant cluster at $ z = 0.41$.
Although the photometric redshift has limitations in the 
case of individual objects, 
we show that the method can be applied successfully to detect an excess 
of objects at a given redshift through a statistical analysis of the 
photometric redshift ranges allowed for the  different objects in the 
field. This is the basic principle of the method we apply on the 
field of Q2345+007.

The outline of the present paper is the following. In the second section,
we briefly summarize the photometric data.
The data reduction and the construction of the photometric catalogue are
discussed in section 3. The method of photometric redshifts is 
introduced and discussed in section 4, together with two 
illustrative examples 
on well-known clusters of galaxies at different redshifts.
In section 5, the population of galaxies responsible
for the excess in number counts at $z \sim 0.75$ is characterized and
its spatial distribution is determined. We also discuss in
section 6 on the existence of other concentrations at $z=0.28$ and 
$z \simeq 1.2$, as well as at the redshifts identified in the absorption 
spectra of the QSO. The SEDs of the arclet-candidates are
studied in section 7. The final discussion and conclusions 
are in section 8. We give the photometry of the double quasar
in appendix.

\begin{table*}
\caption[]{Journal of observations, characteristics of the photometric system 
(filters + detectors) and photometric properties}
\begin{flushleft}
\begin{tabular}{lllrrrrrrrrrrrr}
\hline\noalign{\smallskip}
Date & filter & Detector & Field & Exp. & FWHM &
 $\lambda_{eff} $ & $\Delta\lambda $ & $T_{max}$ &
compl. & limiting & $\mu_{\lambda} (1 \sigma$) \\
  &  &  & $(\arcmin)$ & (sec) & (\arcsec) & (nm) & (nm) & 
 &  mag. & mag. &  mag/$\arcsec^2 $ \\
\noalign{\smallskip}
\hline\noalign{\smallskip}
13-14-15/10/90 & B$_J$ & RCA2 & $2.1 \times 3.4$& 14500 & 1.2 & 447.8 & 142.7 & 0.82 
&  27.5 & 29.0 & 28.7\\
16/10/90 & R & RCA2 & $2.1 \times 3.4$& 3000 & 1.4 & 645.8 & 112.4 & 0.73 
&  26.2 & 26.8 & 26.6 \\
14-15-17/10/90 & I & RCA2 & $2.1 \times 3.4$& 7200 & 1.2 & 812.8 & 126.0 & 0.45
&  25.8 & 26.2 & 26.2 \\
5-7/10/93 & J & NICMOS3 & $2.1 \times 2.1$& 9950 & 1.2 & 1237.0 & 203.1 & 0.94
&  24.5 & 25.5 & 25.4 \\
6/10/93 & K' & NICMOS3 & $2.1 \times 2.1$& 4295 & 1.1 & 2103.2 & 359.5 & 0.96
&  22.8 & 23.5 & 23.1 \\
\noalign{\smallskip}
\hline
\end{tabular}
\end{flushleft}
\end{table*}

\section{Optical and near IR observations}
Data on the field of Q2345+007 come from two different
runs at the 3.6m Canada-France-Hawaii Telescope (CFHT).
Optical images (BRI) were obtained in October 1990 at the prime focus of the
telescope, using the CCD-{\it RCA2} $640 \times 1024$ binned $2 \times 2$, 
with an equivalent pixel size of
$0.41 \arcsec$. Details can be found in Mellier et al. (1994) and
Kneib et al. (1994).
Near-IR images (JK') were obtained at the Cassegrain focus in October 1993,
with the {\it Redeye} camera set in its wide-field mode, 
with a pixel size of $0.50 \arcsec$
(CFH Redeye Users Manual, Simons 1993). The 
details of the observations are summarized in Table 1, as well as the 
characteristics of the photometric system used, taking into
account the response of the different detectors and filters.

All the final images were obtained using the {\it
shift and add} technique developped by Tyson (1988). Near-IR
images were reduced with special care, as explained in next
section. A standard calibration procedure was used in all the
cases. In the optical range, we used the standard stars in the
photometric fields given by Gullixson et al. (1995). A photometric 
sequence exists in the field of Q2345+007 (Tyson \& Seitzer, 1988),
which is especially useful after correction for color effects due
to the difference between the R and I CFHT filters and the Gullixson
et al. filters. In the near-IR, the calibration was done through
the J and K' standard stars given in the Redeye Users Manual (Simons 1993).
We have verified that the magnitude difference K-K' is less than 0.1
magnitudes for all the stars, a value smaller than the absolute 
photometric uncertainty in most cases.
The seeing was reasonably good in all the runs, but the sampling 
of the images is poor. The FWHM given in Table 1 corresponds to the 
final coadded images.

\begin{figure}
\centerline{\psfig{figure=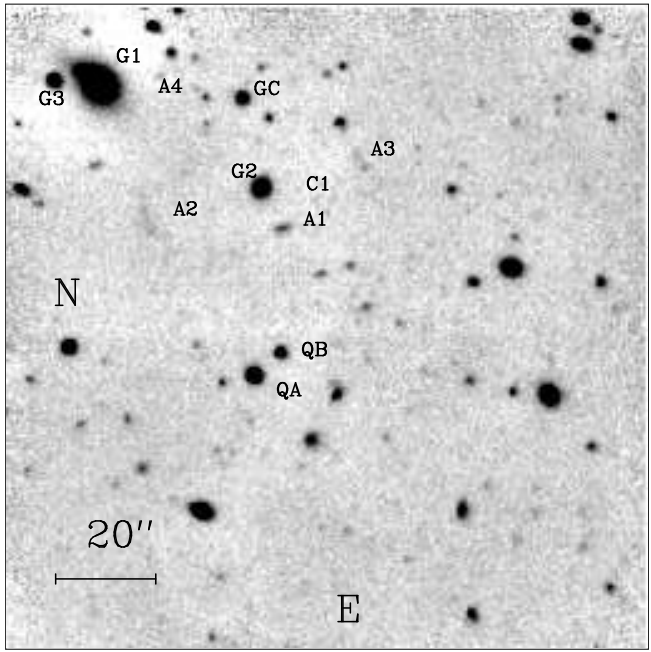,height=8cm}}
\caption [] {J-Band image of the field Q2345+007.
The total exposure time was 9950 sec. The main objects presented in the
paper are identified}
\end{figure}

\begin{figure}
\centerline{\psfig{figure=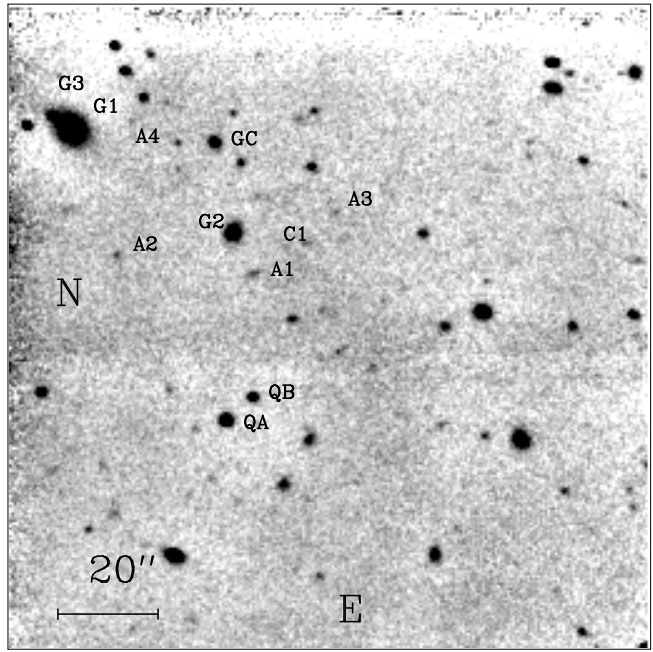,height=8cm}}
\caption [] {K'-Band image of the field Q2345+007.
The total exposure time was 4295 sec. The main objects presented in the
paper are identified}
\end{figure}

\section{Data reduction and photometry}
   Optical images were reduced using the standard IRAF procedures for 
superflatfield and coaddition. In the near-IR, a particular care is necessary
because of the large number of exposures (typical individual
exposures are about 45 seconds in K' and 130 seconds in J) and the sky 
background variations on short time scales, requiring a frequent 
sampling of the sky frame.
As the field of Q2345+007 is almost free of bright objects, it has
been used to obtain a superflatfield, and no comparison empty field
was needed. A discussion about the problems associated with the near-IR
reduction can be found in the Redeye Users Manual. 
Separate superflatfields for images with different sky levels have been
built. In practice, it was necessary to
perform a superflatfield with 10 to 20 successive images. After 
flat-fielding, the images were recentered and coadded using
the IRAF facilities. The resulting J and K' images of the field
are shown in figures 1 and 2 respectivelly, with the identification 
of the main objects presented in this paper. A deep B image of the 
same field can be found in Mellier et al. (1994). The common area of
the optical and the near-IR images is about $2' \times 2'$,
centered on the double quasar, and it largely includes the center of
the shear field detected by Bonnet et al. (1993). 

The photometry was obtained with the AMAPHOT
package developped at the Observatoire Midi-Pyr\'en\'ees. The
detection of objects on the different frames is made at a $1 \sigma$ 
level above the local sky background, with a minimum size requirement
of 4 contiguous pixels above the detection limit. Since the
method presented in this paper is based on the SEDs, only those
objects found on at least two different filters are considered in a 
first step.  Elliptical parameters (centroid, ellipticity and orientation)
are determined for each object from the first and second order moments
of the pixels within its connected domain above the detection isophote.
Isophotal magnitudes are obtained by integrating the profiles computed
along the major axis of the distribution up to a given limiting isophote: 
B$_J$ = 28 mag/arsec$^2$, R = 27 mag/arcsec$^2$, I = 26 mag/arcsec$^2$, 
J = 25 mag/arsec$^2$ and K' = 23 mag/arcsec$^2$, values which are very 
close to the $1 \sigma$ detection limit.
The internal photometric accuracy, obtained by comparing the magnitudes 
coming from different individual images of the same object, is 
better than $\sim 0.05$ magnitude for $B \leq 24$ and $\sim 0.1$ magnitude
for objects with $B \geq 25$. The absolute accuracy is
typically $\sim 0.1$ magnitudes, even for the brightest objects,
when all the usual sources of error are considered (photometric absolute 
calibration, different sampling or seeing conditions), and it increases up to 
$\sim 0.5$ magnitude for the faintest objects of the sample.

We faced specific problems when computing color indices from our data 
because of the wide spectral range covered by the photometry. 
Colors are usually obtained from isophotal magnitudes 
inside a reference isophotal contour applied on all images. But,
in our case, taking a given filter as the reference  
would produce a color-selection effect and could affect the measured color 
distribution. Besides, the morphology and spatial extent 
of an object could be quite different in two extreme filters. 
The important point is to integrate the flux within the same area 
in all the different filters, in order to obtain a SED which is a
good measure of the averaged properties of the object. In order to solve this 
difficulty, we computed an ``average'' set of shape parameters for 
each object as described below: first, we build a sub-image corresponding 
to the average of all images where the object is detected, after 
locally recentering it in the center of the raster and normalizing.
The normalisation consists on setting all the images to the same 
equivalent exposure-time and total transmission.
Secondly, we compute for each object the averaged elliptical 
parameters (center, ellipticity and orientation) from the first and
second order isophotal moments, as well as the mean isophotal radius 
along the major axis, where the limiting isophote 
corresponds to $1 \sigma$ above the local sky value. While the mean
isophotal radius is rather arbitrary, it gives a measure of the averaged 
size of the object and it is used only to compute the color indices.
Some objects with peculiar morphologies, very different from an 
elliptical shape, such as the arclet-candidate A2, require a special 
reduction procedure. In this
case, instead of fitting the shape of the object by an ellipse and to
integrate it inside elliptical annuli, colors and magnitudes are calculated
directly within the (irregular) isophotal limiting contours.

It is also important to verify the reliability of the SEDs derived from
photometry to prevent spurious results due to color effects. In particular,
this could happen when the magnitudes of the
standard stars used for calibration are not {\it exactly} in the same
photometric system as the observations, or when the galactic reddening is
strong. This point is especially important in 
the near-IR where temporal variations of the sky 
transmission can induce color effects in the global response of the system. 
In order to test
the quality of the photometry, we have computed the expected colors for stars
in this photometric system, and we have verified that the observed
sequence of stars on the different color-color planes 
is compatible with the prediction within the errors. The separation of 
stars in the field was obtained with the AMAPHOT package which provides 
a robust test to separate stars from galaxies efficiently up to a magnitude of 
$B_J \sim 23$ (see Pell\'o \& V\'ilchez-G\'omez 1996 for more details). 
Finally, the correction for galactic reddening effects is lower
than the photometric errors, and it was neglected because the galactic 
latitude of the field is high enough. 

The final catalogue contains 849 objects (stars and galaxies) with at least one 
magnitude on the whole field. 559 objects with at least one magnitude are present on the field 
common to all the filters (about $1.9\arcmin \times 1.9\arcmin$). From this sample, 
227 objects are detected in the 5 images, and 550 have at least one color index.
Table 1 summarizes the photometric properties of the catalogue. The limiting 
magnitude defines the faint end of the magnitude distribution which contains
10 \% of the whole sample of objects.

\section{Multicolor analysis of the field}
\subsection{The method}
The aim of our method is to look for an excess of objects at
a given redshift along the line-of-sight. If a cluster is
detectable, it will appear as a peak in the redshift
distribution with respect to an empty field, and will also produce
a significant excess in the 2D distribution of objects. 
In the case of Q2345+007, we searched for any enhancement of 
objects in the redshift distribution first, and then analysed the location 
and center of a possible excess in the 2D distribution of galaxies, as well as
its significance for the gravitational lens system. 
Once the presence of such an excess has been established, 
multicolor photometry can also be used to characterize its population of galaxies.
It is important to determine the 2D distribution and the 
spectrophotometric properties of galaxies to check
if the excess corresponds to a high redshift cluster.

The first step in the method is to determine a photometric redshift range 
for each object from its SED, derived through BRIJK' photometry. The SED is
equivalent to a very low resolution spectrum in a large
wavelength range, from 0.35 to 2.4 $\mu m$. 
Photometric redshifts are determined as the intersection of the
``permitted'' redshift domains coming from different color-redshift
diagrams, which are calculated using the Bruzual's code
for the spectrophotometric evolution of galaxies (Bruzual \& Charlot,
1993), taking into account the transmission functions of the
photometric system. The IMF is that of Miller \& Scalo (1979), with a
lower and upper cutoff masses of 0.1 and 125.0 $M_{\odot}$ 
respectively. Throughout this paper, we assume $H_0=50$ 
km $s^{-1}$ $Mpc^{-1}$ and $q_0=0.1$. The present age of galaxies is 
assumed to be 15 Gyr, which corresponds to a redshift for galaxy 
formation of $z_f = 5.3$. The results obtained are almost independent
from the choice of the IMF and the present age assumed for galaxies,
provided that present-day ellipticals are older than 9-10 Gyr. For our
purposes, the main influence of $H_0$ comes from the maximum age allowed
to present-day ellipticals. It is easy to show that the permitted redshift
domains become narrower when the spectral range is increased to the
near-IR, so the individual photometric redshifts become more 
accurate in this case. 

In practice, we used as many independent colors as possible to constrain 
the photometric redshift of a given object, namely the 4 colors 
B$_J$--R, R--I, I--J and J--K' for objects detected in all the filters. For 
each color, a maximum 
(red) and a minimum (blue) curves are calculated in the color-redshift plane, 
which define the permitted area for any galaxy. The reddest curve is that 
of an old single burst of star formation, such as an old elliptical galaxy
(hereafter E), and the bluest curve is that of a continuous star 
forming system, such as an irregular magellanic galaxy (hereafter Im). The use 
of filter bands from B to K' allows to obtain simultaneously an optimal fit 
of short time scale phenomena (such as a burst of star formation) as well as
the averaged behaviour of the system (as given by the old population
of stars seen in the J and K' filters). Each observed color is compared 
to the predicted color-z diagram, taking into account the photometric 
errors, and the permitted redshift range is obtained. The code used to compute 
models does not include metallicity effects on the stellar
population or internal absorption effects. Nevertheless, such effects
are expected to be of second order in the determination of the redshift. 
At a given z, any ``normal'' galaxy fits between the two extreme models if 
its spectromorphological
type is let completely free. This is almost independent from the existence 
of emission-lines, because the SED built from broad band filters is generally 
dominated by the continuum and absorption features (the 4000 \AA \ and
Balmer breaks in the visible). Obviously,
when we are dealing with an object whose SED is different from
a normal galaxy, no photometric redshift is found. This is the case
of the double quasar, for example. The final redshift range for an
individual object will be narrow (about 0.1, to take into account
the precision in the color-redshift diagrams in most cases) or wide
depending on the number of colors available, the width and the sampling
properties of the filters used and the photometric accuracy. As the 
color calibration accuracy has been tested on a sample of stars, the reliability of 
the SEDs essentially depends on the individual photometric 
uncertainties. In summary, even if the redshift 
range is wide for some individual objects, the existence of a cluster can be 
evidenced statistically because most objects belonging to it will be compatible 
with its redshift. We discuss this point below, in the case of two well known 
clusters of galaxies, namely A2218 and A370, where similar optical and 
near-IR deep photometry is available and where the result concerning 
the redshift is unambiguous. 

\subsection{A test on the clusters of galaxies A2218 and A370}
   In the case of A2218, a set of optical images is available in 
filters B, g, r and z, as well as near-IR images in the same filters 
as Q2345+007. Details concerning the photometry of this cluster
can be found in Le Borgne et al (1992), Pell\'o et al. (1992) and
Kneib et al. (1995). The mean spectroscopic redshift is
$z = 0.1756 \pm 0.0068 $ (Le Borgne et al. 1992). The uncertainty 
on the redshift is intrinsic, due to the velocity dispersion of the 
cluster (1370km/s) and does not correspond to observational errors.
A catalogue of 232 objects was obtained in the common
field of all the filters, excluding objects identified 
unambiguously as stars. The automatic procedure has given a
photometric redshift interval for 182 of them ($78 \%$ of the sample). 
Figure 3a 
shows the redshift distribution, where all the permitted redshift 
intervals have been cumulated with a weight of 1.
A significant excess of galaxies at a mean redshift of $z = 0.17 \pm 0.05$
can be seen. The error bar is a measure of the FWHM of the peak
around the mean value. This result is fully compatible with the 
spectroscopic value. The photometric redshift matches perfectly 
the spectroscopic result (Le Borgne et al. 1992)
for 85 \% of the objects for which J and K' photometry is available.
This is the case of the arcs at $z_{sp} = 1.034$ and 
$z_{sp} = 0.702$, for which the automatic procedure has found
a photometric redshift between 0.825 and 1.1 for the former and
between 0.650 and 0.725 for the later. A significant secondary and wider 
peak exists around $z \sim 0.6$, due to the distribution of the 
background population of galaxies. A significant 
contribution is expected from the many bright arclets observed in this field.
No comparison blank field, obtained in the same conditions and
filters as the cluster, is available in this case to properly estimate
the field contamination, but the contrast cluster-field 
is high enough to detect the excess.

\begin{figure*}
\centerline{\hbox{ \psfig{figure=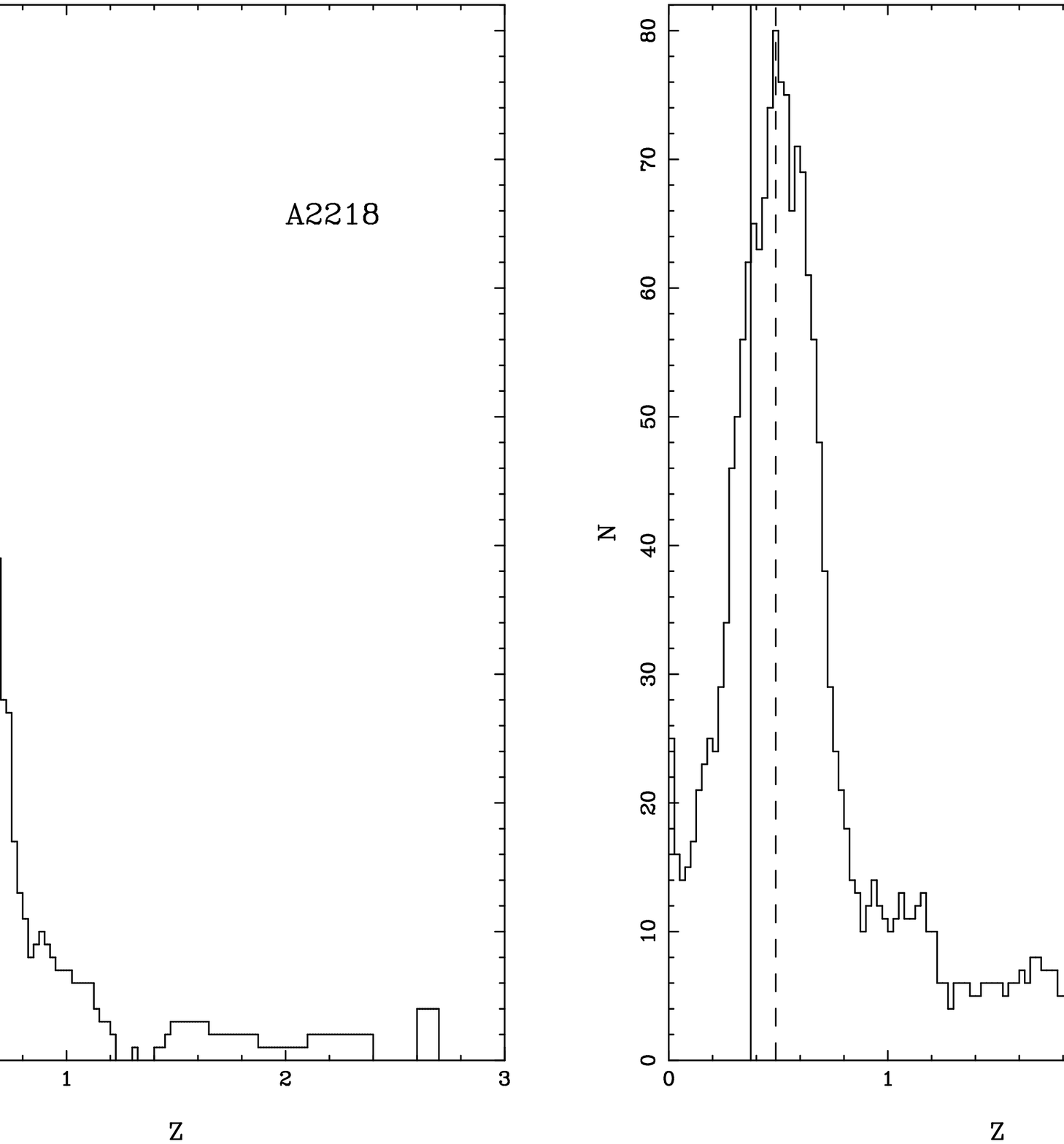,height=9cm}
\psfig{figure=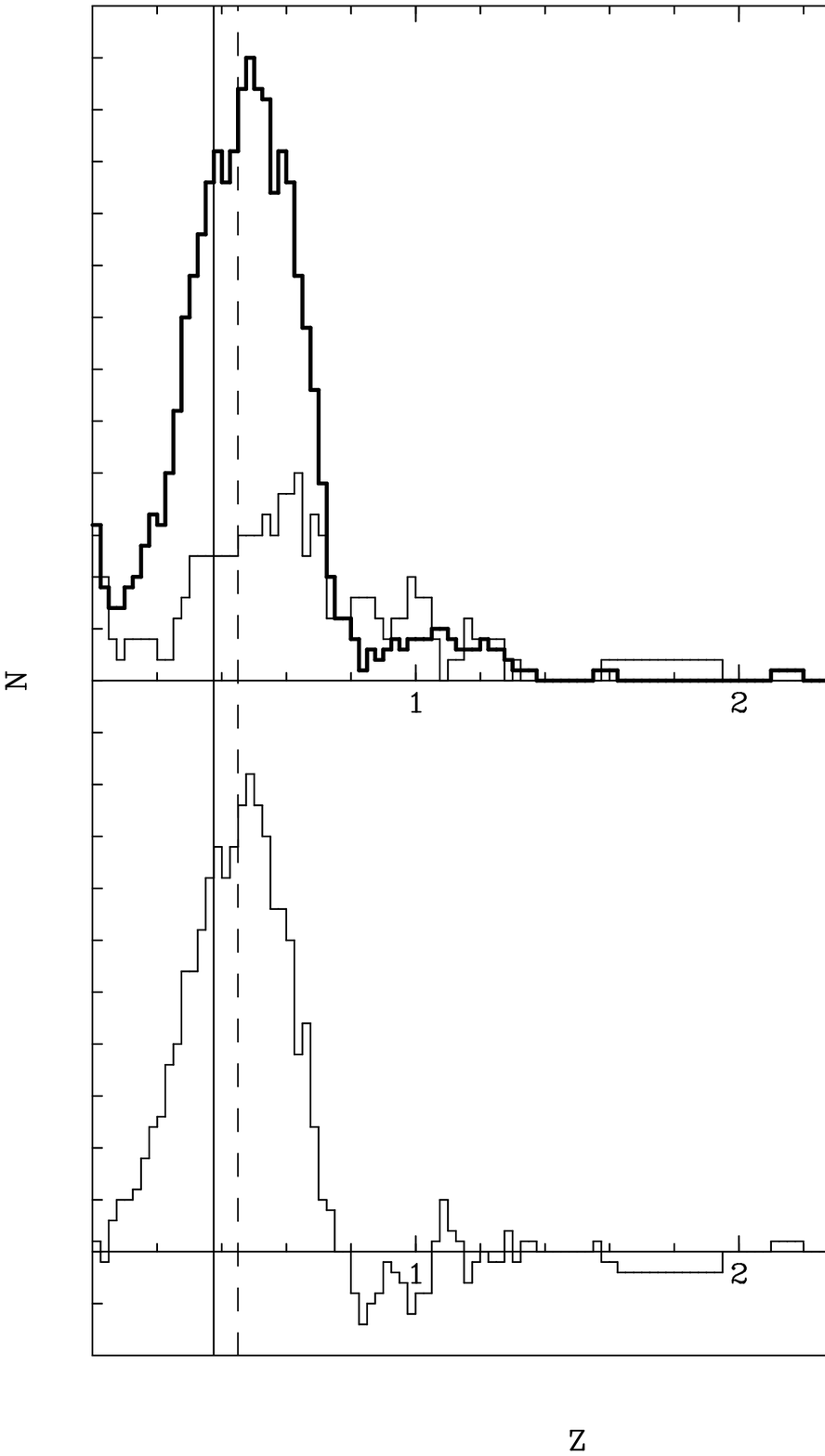,height=9cm}}}
\caption [] {Photometric redshift distribution of objects in different fields: a) A2218 
(raw data); b) A370 (raw data); c) A370 compared to a blank reference field, 
after normalisation, with the net distribution in the bottom.
The spectroscopic and the mean photometric 
redshifts are indicated by solid and dashed lines respectively 
}
\end{figure*}

In the case of A370, the photometry
is available in the same filters and has been obtained during the same runs
and under the same conditions as for Q2345+007. The mean spectroscopic redshift 
is $z=0.374 \pm 0.001 $ (see Mellier et al., 1988, and Soucail et al., 1988,
for more details about the spectroscopic survey on this cluster).
The catalogue of the field common to all the filters
contains 345 objects, and the automatic procedure has given 
a photometric redshift interval for 252 of them ($75 \%$ of the sample).
The raw redshift distribution, obtained in the same way as in A2218, is
shown in figure 3b. The excess of galaxies is at a 
mean redshift of $z = 0.48 \pm 0.11$, so it is roughly in agreement 
with the spectroscopic value, although slightly higher, and 
the peak obtained is wide. The comparison 
blank region defined below in Sect. 5.1 (field 2) was used to estimate
the field contribution by subtraction. The redshift distributions in the cluster
and the blank fields are shown in figure 3c, after correction for the different
sizes of the two regions. The two catalogues used to compute the redshift 
distributions have been limited within the magnitude of completeness in 
{\it all} the filters and, when this magnitude is different 
in the two fields, the limit is given by the brightest of them. The
final catalogues contain 193 and 50 objects respectively in the cluster 
and the blank field, and about $80 \%$ of them have a redshift interval
assigned. The net excess of galaxies in the field of A370 is then at a 
mean redshift of $z = 0.45 \pm 0.08$. 80 \% of the objects for wich a 
spectroscopic redshift exists (according to Soucail et al., 1988) 
are correctly identified photometrically. This is the case of the 
giant arc at $z_{sp} = 0.724$, for which the automatic procedure 
has found a photometric redshift $z = 0.50-0.75$. 

   The maximum peak obtained in A2218 is quite narrow compared to that found
in A370 and there are two reasons for this. First, the redshift of A2218 
is low, so the photometry is more accurate because the objects are 
brighter in average and the SEDs show a 
low internal dispersion in redshift and spectromorphological types. Secondly,
the more important spectral feature in the visible range at such intermediate 
redshifts is the $4000 \AA$ discontinuity, and many narrow optical filters are available 
in the field of A2218 to map the SEDs around this wavelength region. 
Figure 4 shows the difference between the spectroscopic and the photometric 
redshift as a function of the magnitude for different objects in the fields of A2218
and A370. There is no clear evidence for an increase of errors as a function of the
magnitude. It is worth noting 
that the contrast between the cluster and the field increases when the near-IR photometry is 
available, and this effect is mainly due to the narrowing in the individual
permitted redshift ranges. The mean photometric redshift obtained for A370 is higher than
the spectroscopic value and the fit is worse compared to A2218. A straightforward
simulation shows that such a systematic bias 
towards higher redshifts is expected for E-type galaxies at $0.2 \leq z_{sp} \leq 0.4$ 
when the filters BRIJK' are used, and it disappears beyond  $z_{sp} \geq 0.5$. 
The effects of sampling along the SEDs on the inferred
redshift distribution will be studied in details in a forthcoming 
paper (Miralles et al., in preparation).

\begin{figure}
\centerline{\psfig{figure=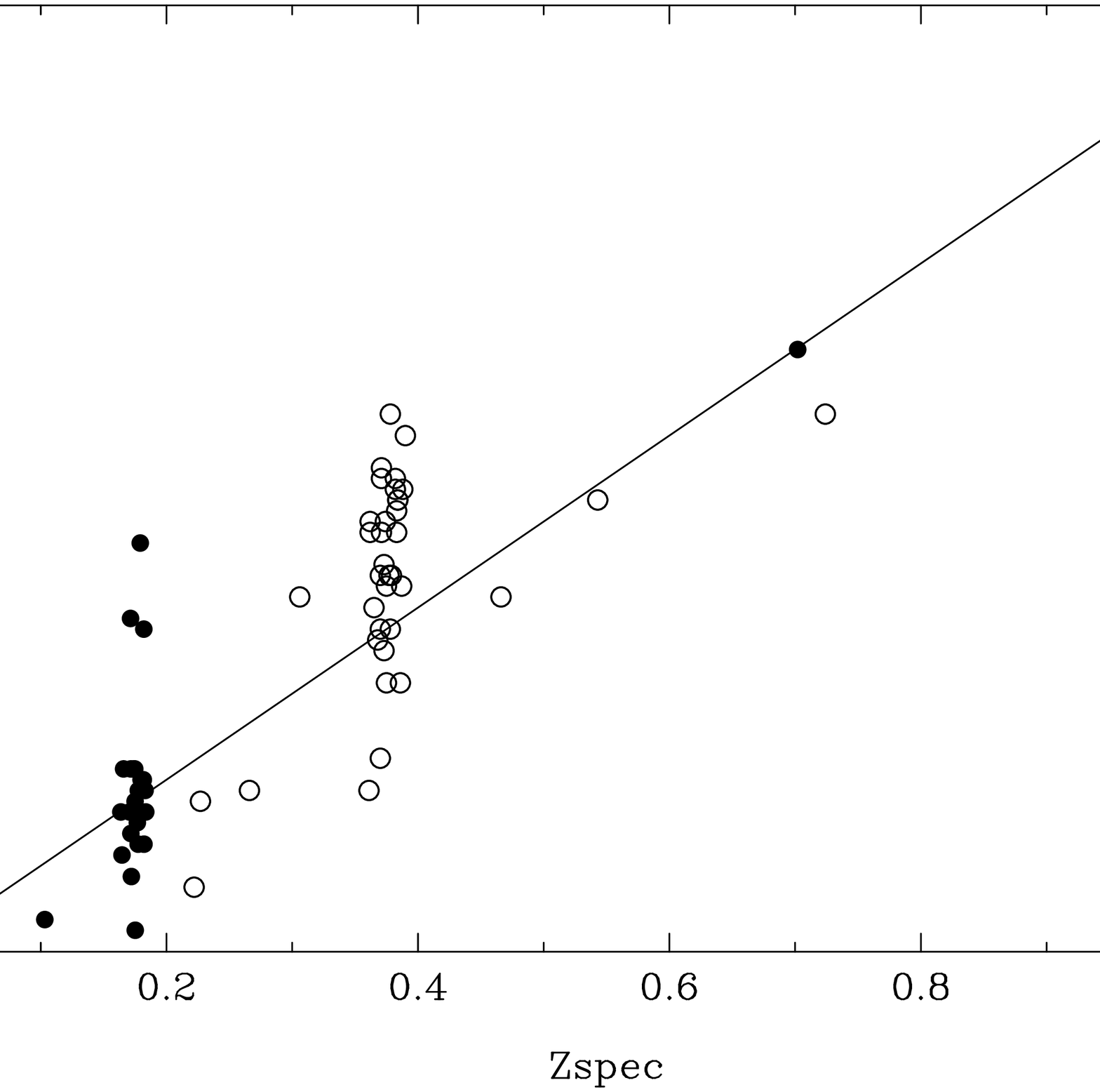,height=7.5cm}}
\centerline{\psfig{figure=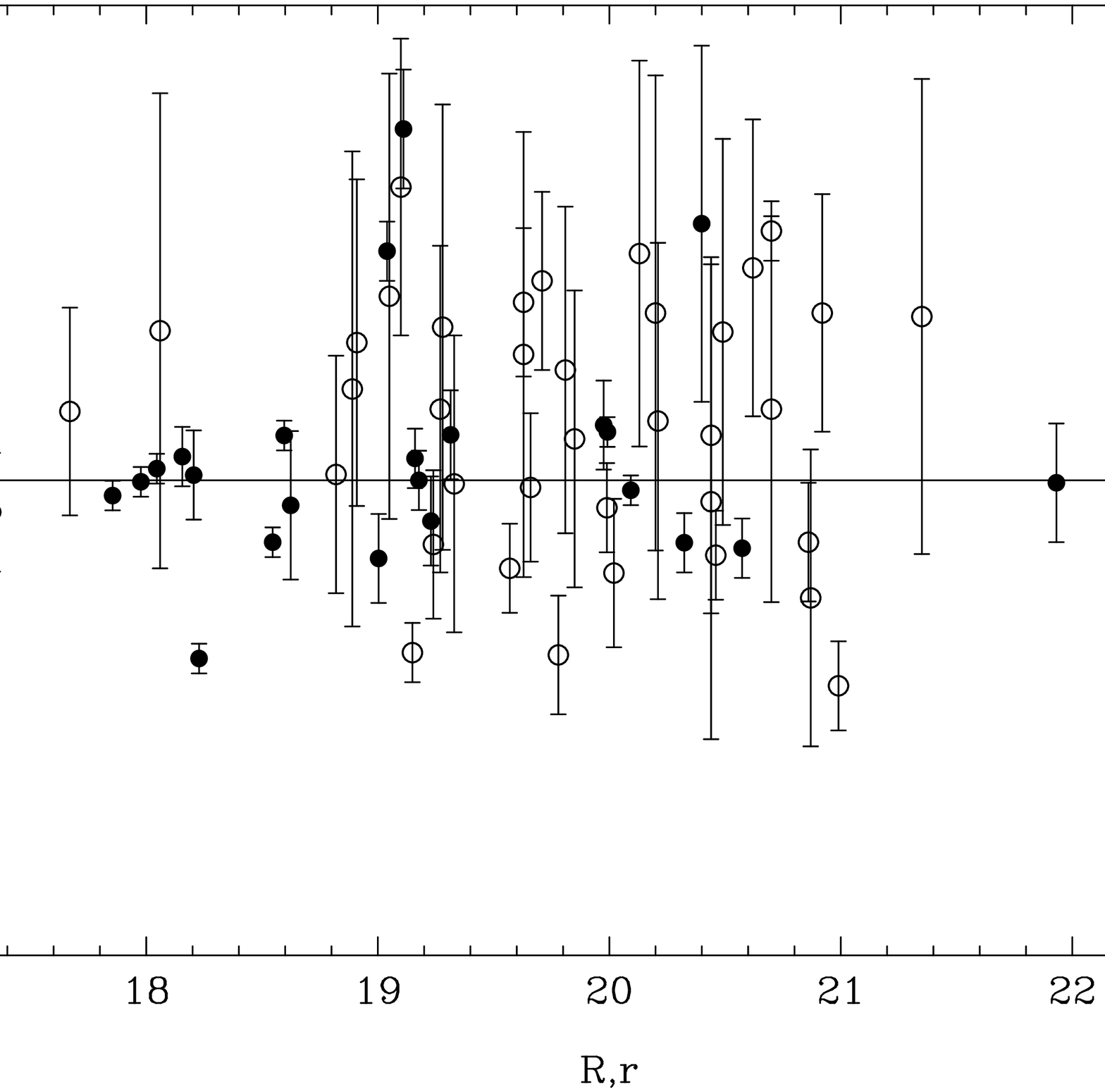,height=7.5cm}}
\caption [] {a) Photometric redshift versus spectroscopic redshift
in the fields of A2218 (full dots) and A370 (open dots).
b) Difference between the spectroscopic and the photometric redshifts
as a function of the magnitude for objects in the fields of A2218 (full dots) and 
A370 (open dots). Error bars show the permitted photometric-redshift intervals. 
Magnitudes are given in r for A2218 and in R for A370.
}
\end{figure}

\section{Detection and characterisation of a cluster of galaxies at z $\sim$ 0.75}
\subsection{Detection of excesses of galaxies at different redshift planes}
The method presented above was applied to the catalogue of 550
objects with at least one color on the common field of Q2345+007. The 
automatic procedure enabled to calculate a permitted redshift range for 
$78\%$ of them, and the cumulative histogramme for the redshift
distribution has been obtained using the same method as in Sect. 4.
The contrast of a possible high redshift cluster with
respect to the normal distribution for a blank field is expected to be lower
than for A2218 or A370, because the field contamination strongly increases
at faint magnitudes. In order to estimate this contamination,
two regions of about the same surface have been defined by dividing the 
whole field in two halves: one contains the 
double QSO and the excess of galaxies already detected by Mellier 
et al. (1994) (field 1), and the other one is considered as an 
``empty'' comparison region (field 2). The completeness magnitudes in the fields
1 and 2 are the same in all the filters. 
We can check on the excess of galaxies in the field of the double QSO
through galaxy counts. Even if the contrast between field 1 and the blank is
expected to be low, it is interesting to compare the present results
with those found previously by other authors. Table 2 gives the results on the B$_J$ 
and K' galaxy counts in the two fields, compared to those already 
obtained by Tyson (1988) in B$_J$, and Djorgovski et al. (1995) 
and Gardner et al. (1993) in K'. 
Our results have not been corrected for incompleteness, so they only
compare to those found in the literature up to the completeness magnitudes.
In general, the present results are in agreement with those
found previously within the errors. 
There is an excess in K' compared to Djorgovski et 
al. (1995) at the faintest magnitudes, in the two fields, but there is
no clear evidence for an excess in field 1 compared to the blank. 

The superposition of the raw 
distributions in redshift in the two regions are shown in figure 5a, 
and the net distribution in the field 1, obtained by subtraction, is given 
in figure 5b. A clear 
narrow peak appears at $z = 0.75 \pm 0.08$ and a wider one at $z=1.2 \pm 0.1$,
as well as a secondary wide peak at $z \sim 1.8$. All these peaks 
correspond to excesses of galaxies in the 2D distribution, with different 
levels of significance, as explained below.
The first one is identified as a cluster at $z \sim 0.75$,
as it is discussed below. The redshift of this excess is fully compatible 
with the absorption 
system detected in the spectrum of the B component of the quasar, 
a first evidence in favour of a lens at such a redshift. 
The second peak
at $z \sim 1.2$ does not correspond to any absorption system of the quasar.
The third peak centered at $z \sim 1.8$ is the widest and 
the less significant (the maximum contrast in the redshift distribution 
is about $3 \sigma$), but it contains 
the other absorption systems identified by Steidel and Sargent (1991) in the 
spectra of the QSOs, between $z=1.49$ to  $z=1.98$. Most galaxies compatible 
with the two high-redshift peaks lie beyond the completeness limit in at 
least one filter, as it is shown in figure 5b, hence the conclusions concerning
their 2D distribution have to be taken with caution. More detailed results concerning 
the two higher redshift-strips can be found in next section.  

\begin{figure}
\centerline{\psfig{figure=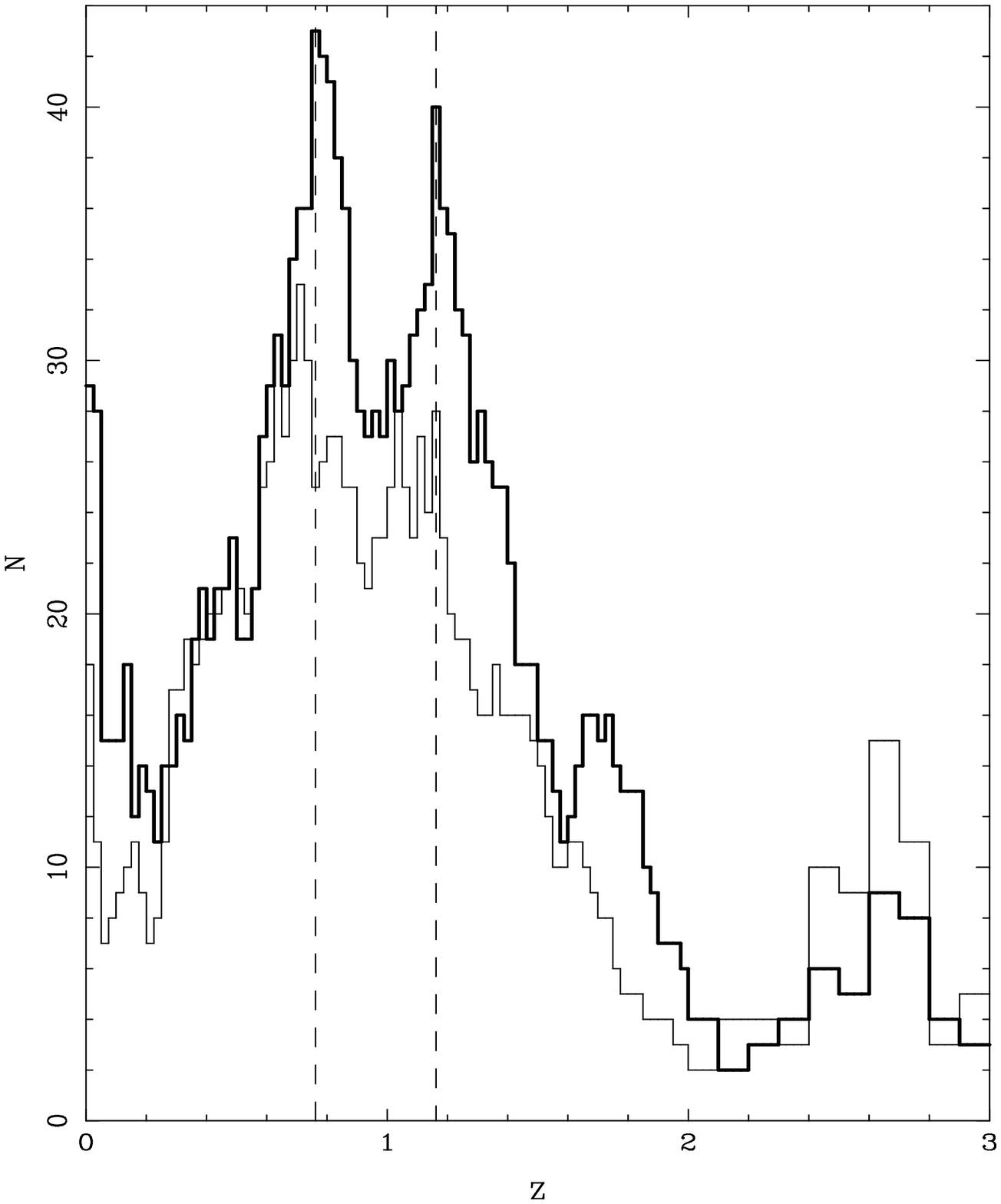,height=9cm}}
\centerline{\psfig{figure=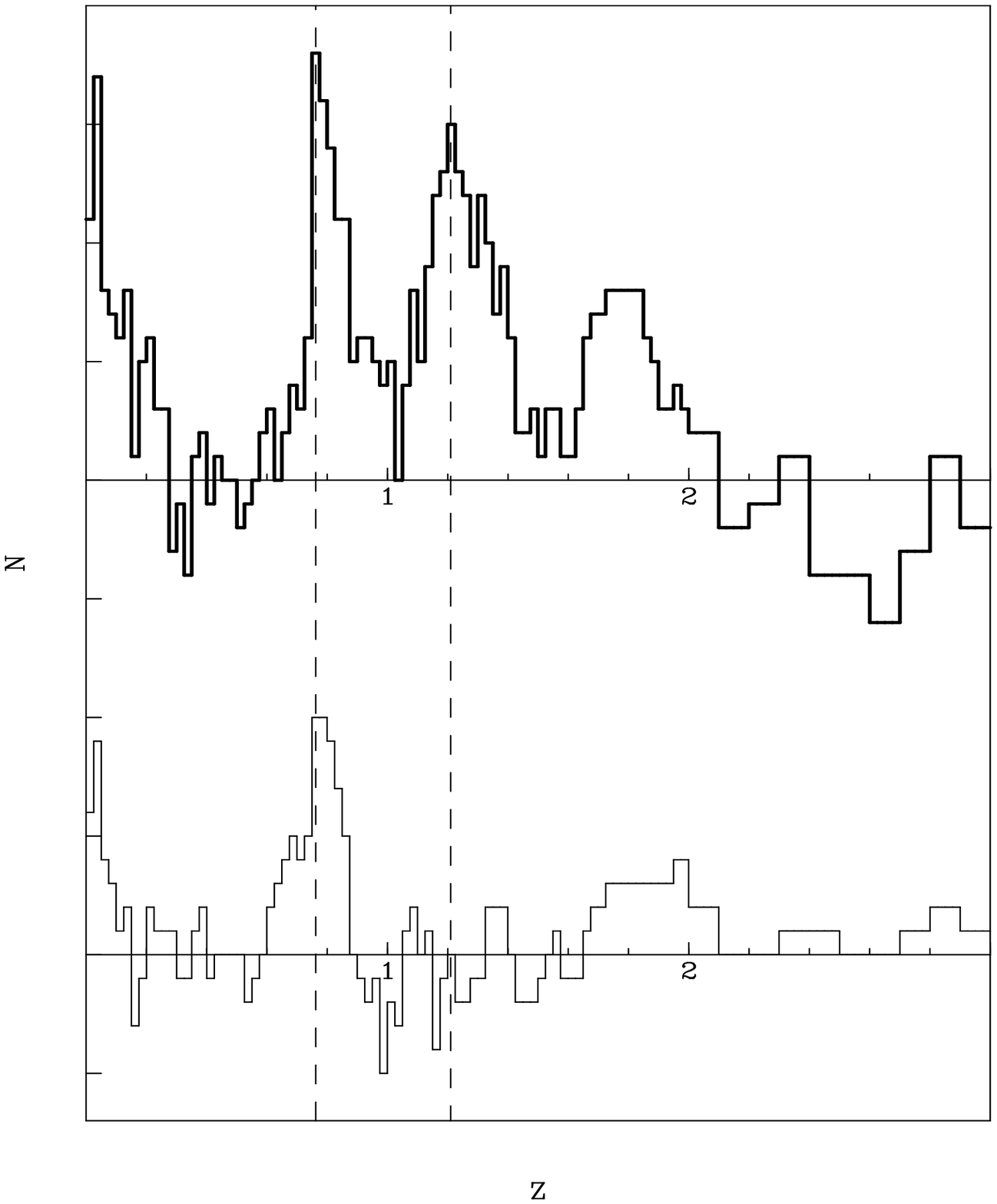,height=9cm}}
\caption [] {Redshift distribution of the objects in the field of
Q2345+007: a) raw data; the
thick line correspond to the region of the field containing the double
QSO and the excess of galaxies already detected by Mellier et al. (1994)
(field 1),
whereas the distribution for the blank field (field 2) is shown for
comparison by a thin line; b) net redshift distribution in the field 1,
for all the sample (top) and for objects 
brighter than the completeness limit in B$_J$ (bottom).
}
\end{figure}
   
\begin{table*}
\caption[]{Galaxy counts in the field of Q2345+007 in B$_J$ and K', 
compared to those obtained by Tyson (1988) in B$_J$ and Djorgovski 
et al. (1995) in K'. Fields 1 and 2 correspond to that of the double
QSO and the comparison field respectively. The field {\it all} means
the whole field covered by the image in the given filter, and not the 
common field (field 1 $+$ field 2). Magnitude intervals are one 
magnitude wide, centered on the value. 
}
\begin{flushleft}
\begin{tabular}{lrrrrrrrrr}
\hline\noalign{\smallskip}
Filter B$_J$ & & & & & & & & \\
Mag & N & Log N/mag/deg$^2$ & N & Log N/mag/deg$^2$ & N & Log N/mag/deg$^2$
&  Log N/mag/deg$^2$  & \\
\hline
 & & all & & field 1 & & field 2 & Tyson (1988) & \\
\hline\noalign{\smallskip}
22.50 & 8 & 3.57$\pm$0.15 & 2 & 3.54$\pm$0.31 & 2 & 3.57$\pm$0.31 & 3.6 & \\
23.50 & 27 & 4.10$\pm$0.08 & 4 & 3.84$\pm$0.22 & 4 & 3.87$\pm$0.22 & 4.10 & \\
24.50 & 57 & 4.42$\pm$0.06 & 17 & 4.46$\pm$0.10 & 12 & 4.35$\pm$0.12 & 4.55 &  \\
25.50 & 102 & 4.68$\pm$0.04 & 37 & 4.80$\pm$0.07 & 21 & 4.59$\pm$0.09 & 4.70&  \\
26.50 &  90 & 4.62$\pm$0.05 & 22 & 4.58$\pm$0.09 & 32 & 4.77$\pm$0.08 & & \\
27.50 & 205 & 4.98$\pm$0.03 & 71 & 5.09$\pm$0.05 & 61 & 5.05$\pm$0.06 & & \\
\noalign{\smallskip}\hline
\hline\noalign{\smallskip}
Filter K' & & & & & & & & \\
Mag & N & Log N/mag/deg$^2$ & N & Log N/mag/deg$^2$ & N & Log N/mag/deg$^2$
& Log N/mag/deg$^2$  & \\
\hline
 & & all & & field 1 & & field 2 &(1) Gardner et al. (1993) & \\
 & &  & &  & & &(2) Djorgovski et al. (1995) & \\
\hline
 18.50 & 14 & 4.10$\pm$0.12 & 6 & 4.01$\pm$0.18 & 7 & 4.11$\pm$0.16 & 
3.97$\pm$0.07 (1) &\\
 19.50 & 22 & 4.30$\pm$0.09 & 5 & 3.93$\pm$0.19 & 16 & 4.47$\pm$0.11 & 
4.25$\pm$0.06 (1) & \\
 20.50 & 38 & 4.53$\pm$0.07 & 15 & 4.41$\pm$0.11 & 20 & 4.57$\pm$0.10 &
4.52$\pm$0.13 (2) & \\
 21.50 & 47 & 4.63$\pm$0.06 & 28 & 4.68$\pm$0.08 & 19 & 4.55$\pm$0.10 &
4.60$\pm$0.12 (2) & \\
 22.50 & 166 & 5.17$\pm$0.03& 82 &5.15$\pm$0.05 & 82 & 5.18$\pm$0.05 &
 4.91$\pm$0.08 (2) & \\
\noalign{\smallskip}
\hline
\end{tabular}
\end{flushleft}
\end{table*}

\subsection{Structural properties of the cluster-lens}
   The next step is to study the 2D distribution of objects 
with a permitted redshift interval which contains $z = 0.75 \pm 0.05$. The projected 
number and luminosity densities of such objects were computed through a 
method similar to that proposed by Dressler (1980). Figure 6 shows the isocontour
plot corresponding to the distribution in projected number-density, where a clear 
maximum appears close to galaxy G2. The mean density of objects compatible with 
$z \simeq 0.75 $ is 13.6 galaxies/arcmin$^{2}$ up to the completeness magnitude 
in B$_J$, and the maximum value reaches 4.1 times the mean density. The
contrast in density, defined as $[\rho - <\rho>] / \sigma$, where $\sigma =
<\rho>^{1/2}$, is 11.6. The size of the region where 
the density is higher than 2.5 times the mean value is about $40\arcsec$
($375 h^{-1}_{50}$ kpc at $z=0.75$). The total number of 
galaxies compatible with $z = 0.75\pm 0.05$ within a circular region of
this diameter is 14, whereas only 6 were expected according to the mean 
density of such galaxies over the whole field (field 1 $+$ field 2), so the contrast 
is about $3.3 \sigma$. The results in number-density are similar when we consider 
objects within the completeness magnitude in J. The maximum in projected number-density 
given by the Dressler's method lies as expected between the two clumps reported 
by Mellier et al. (1994, Fig. 1), and its distance to the center of the weak-shear 
field is about $11 \arcsec$ West (see figure 6). The location of the projected
number-density peak is also in good agreement with the excess detected by
Fischer et al. (1994), and
it is also compatible with the position of the lens responsible for the
weak-shear field found by Bonnet et al. (1993).

   The 2D distributions in luminosity-density in B$_J$ and J are also very
similar for the sample of galaxies within $z=0.75 \pm 0.05$. The position of the 
maximum is the same in both filters, and it lies 
less than $5 \arcsec$ northwards from the maximum in projected number-density.
The maximum values in B$_J$ and J up to the completeness magnitudes are,
respectively, 4.6 and 4.1 times the mean luminosity-densities in these filters, 
the mean values beeing 
$3.41 \times 10^{10} h^{-2}_{50} L_{\odot}$ / arcmin$^2$ in B$_J$ and
$7.47 \times 10^{11} h^{-2}_{50} L_{\odot}$ / arcmin$^2$ in J at $z=0.75$
(raw data, no k-correction is applied). A final remark afterwards is that 
the size of the whole field studied (field 1 + field 2) is about 1 Mpc wide 
at $z=0.75$, so field 2 is not expected to be 
an empty region at such a redshift, and the contrast found is then a lower limit. 
In fact, an excess at this redshift is already visible in the raw data 
presented in figure 5a.

\begin{figure}
\centerline{\psfig{figure=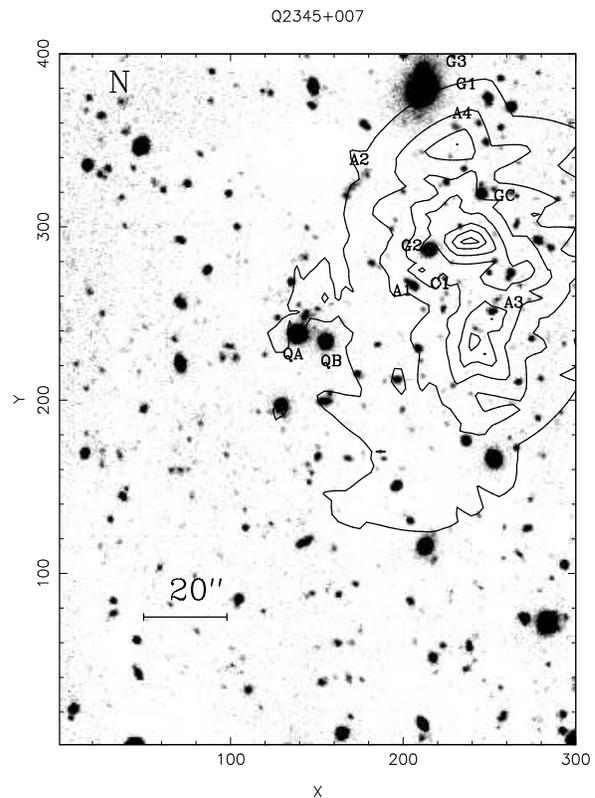,height=11.6cm}}
\caption [] {Isocontour plot of the projected number density of galaxies
compatible with the excess at $z \sim 0.75$, superimposed on a composite
image of the common field. Successive contours correspond to
0.5 times the mean value over the whole field (field 1 $+$ field 2), 
starting at 1. The composite image was obtained by coadding all the
final images in all the filters, after normalization. The main objects 
are identified. C1 is the center of the shear field detected by Bonnet 
et al. (1993). North is towards the upper left corner, and the scale is 
in pixels (1 pixel = $0.41 \arcsec$). 
}
\end{figure}

\subsection{Photometric properties of the cluster-lens}
There are 80 objects detected on the whole field (field 1 $+$ field 2)
within a permitted redshift interval which contains $z=0.75 \pm 0.05$. 
The distribution of all these galaxies on 
color-magnitude and color-color diagrams are presented in figures 7a and 
7b, respectively. Galaxies identified as belonging to other possible redshift systems
are also plotted for comparison. The look-back-time at $z \sim 0.75$ is 8.15 Gyr,
so galaxies are 6.85 Gyr old. A detailed discussion about the dominant stellar
population in the two extreme star-forming systems used can be found in
Bruzual \& Charlot (1993). At $z \sim 0.75$, the main spectral 
absorption feature expected in the visible range is the 4000 \AA \ discontinuity,
which lies between the central wavelengths of the filters R and I. Nevertheless,
these filters are so broad that the R-I color index alone is unable
to discriminate between foreground (beyond $z \sim 0.3$) and background 
objects (up to $z \sim 1.3$), when the spectromorphological type of the galaxies
is let completely free. 
The determination of the redshift has been possible only because we
have strong constrains on the whole continuum from B$_J$ to K' for most objects in the
field. 

   Concerning the general evolutionary trends, the filter B$_J$ is 
very sensitive to short-scale star-forming phenomena. The J band maps the restframe 
from $\lambda \sim$ 6000 to 8200 \AA, so it is very sensitive to the present
main-sequence (MS) population of stars, regardless of the star-formation history 
and for a relatively wide range of redshifts around 0.75. A similar comment can be 
applied to the I band, which samples the restframe wavelength just after the 
4000 \AA \ break: it is also
dominated by MS stars but it is more sensitive to the redshift because, beyond 
$z \sim 0.80$, the 4000 \AA \ break lies inside the filter. Then, with 
some cautions concerning the redshift and the metallicity, the color I-J can 
help in constraining the age of the dominant stellar population. On the contrary, 
the K' band samples the restframe at $\lambda \geq 1 \mu m$, which is dominated by 
red giant or core He burning low-mass stars (depending on the past star-formation 
history), so it is almost insensitive to the present star-formation rate and it
puts some constraints on the past star-forming history. The results on the 
sample at $z \sim 0.75$ evidence that most objects have been undergoing an active 
star-formation activity in a recent epoch,
because they show R-I and I-J colors compatible with those of a continuous star-forming
system, or with those of a young burst of
star-formation (younger than 3 Gyr in most cases). According to their B$_J$-R,
about 50\% of them seem to be continuous star-forming systems, whereas 
the others are more likely young burst systems where the star-formation 
stopped 1 to 3 Gyr ago. Only two objects have red SEDs in these filter-bands, 
compatible with those of an old E-type system. Moreover, about 50\% of these 
star-forming objects are red in the J-K' distribution and show an enhanced 
flux in the K' band compared to what is expected for a single young burst
of star-formation. Then, all these results point to the existence of a generalized and 
recent burst of star-formation activity, less than 3 Gyr old at $z \sim 0.75$, 
on galaxies which already have an old population of stars. We have checked on 
possible bias in the color-distribution of this population of galaxies due to 
the different completeness and limiting magnitudes in the various filters, but 
the result is negative. More than 50\% of galaxies belonging to this sample 
are within the completeness magnitudes in {\it all} the filters.

\begin{figure*}
\centerline{\psfig{figure=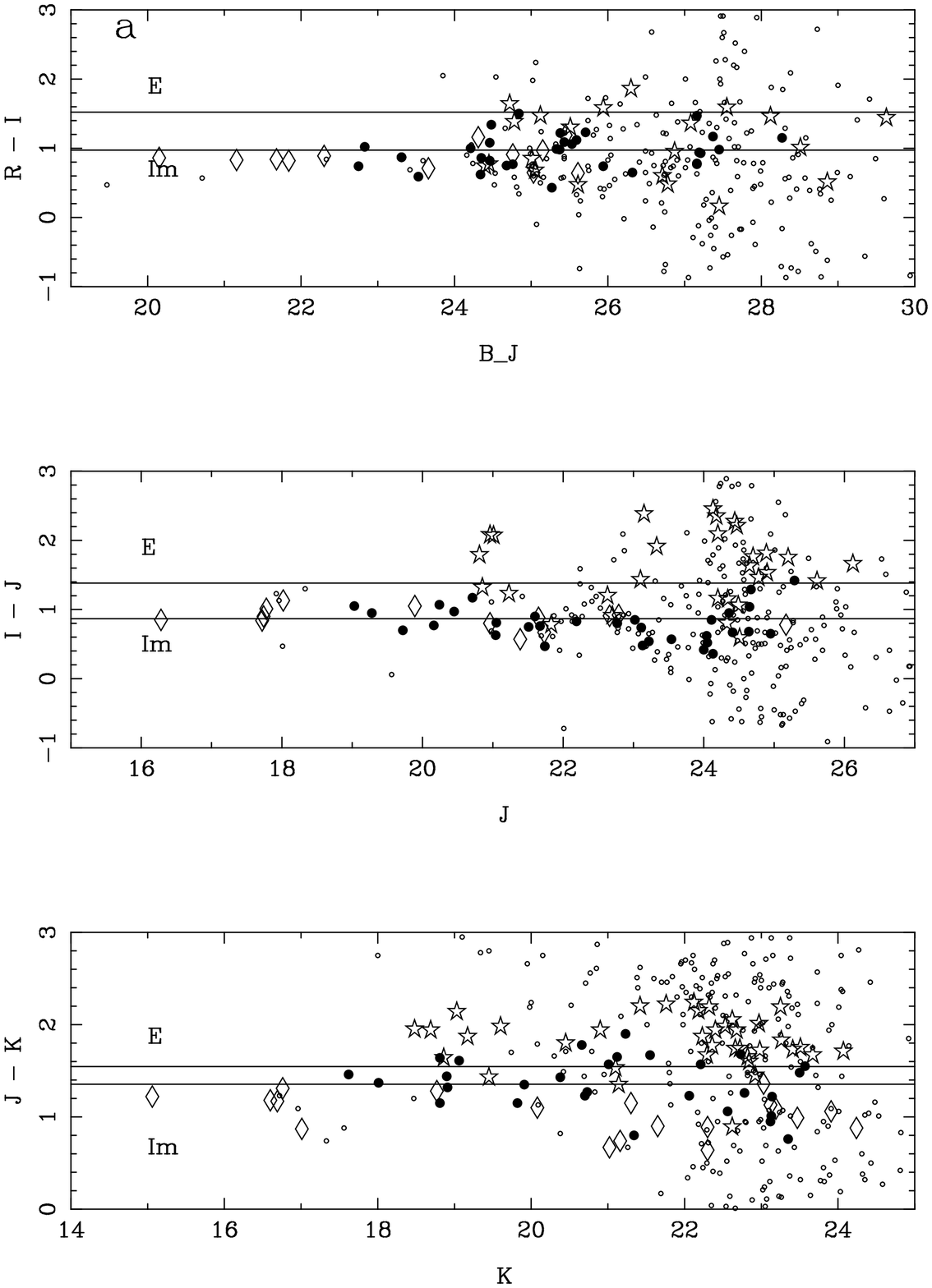,height=11.6cm}
\psfig{figure=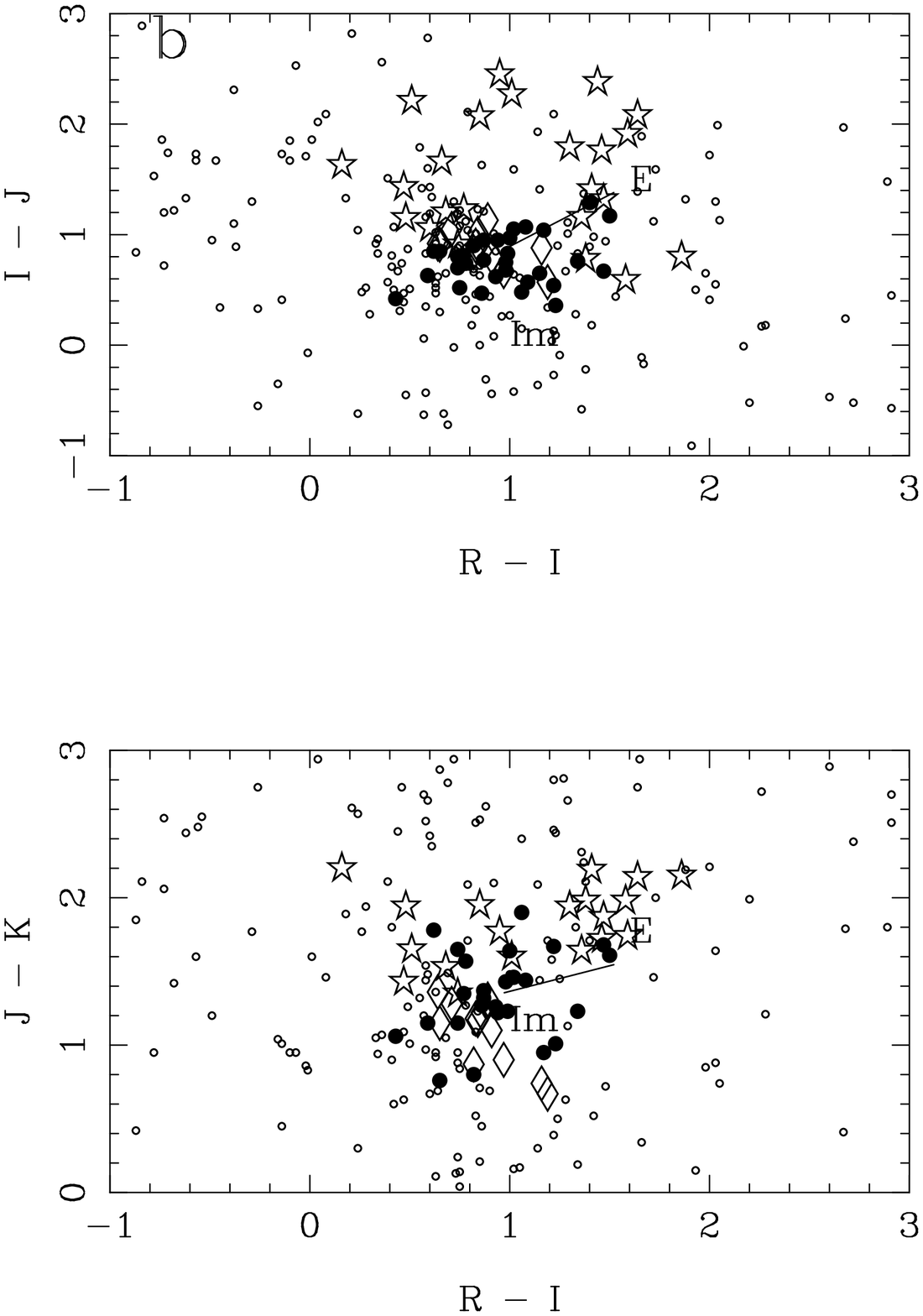,height=11.6cm}}
\caption [] { Color-magnitude (a) and color-color (b) diagrams for objects in 
the field of Q2345+007. Objects compatible with a redshift of $z=0.75 \pm 0.05$
are represented by full dots, whereas objects compatible with $z=0.28\pm 0.05$
and $z=1.2 \pm 0.10$ are drawn as diamonds and stars, respectively. The colors 
predicted for the two extreme types of galaxies, E and Im, at $z=0.75$ are also
shown}
\end{figure*}

   The brightest galaxy in the sample at $z=0.75 \pm 0.05$ is located only 
$13 \arcsec$ away from the position of the 
maximum in projected number and luminosity-density maps in B$_J$ and J.
This galaxy is then proposed as the central cluster-galaxy. Its isophotal B$_J$ 
magnitude is 22.83, so $M_B = - 21.67$ in the rest frame (K-correction
is computed according to its SED), and the corresponding isophotal radius
is about $3\farcs3$ (about $31 h^{-1}_{50}$ kpc). These values
are quite compatible with those expected and observed for the brightest 
cluster-members in more nearby clusters. We have 
used a minimization procedure to fit the best synthetic spectrum coming from a 
single population to its photometric SED. The best fit is obtained 
for a young burst of star-formation 1.3 Gyr old, and the best fit for the redshift
is $z \sim 0.70$, as it is shown in figure 8. The signature of an old stellar 
population appears as an excess in the J and K' mean fluxes. Five
other bright galaxies compatible with $z=0.75 \pm 0.05$ are present within 
a distance of $30''$ (about $280 h^{-1}_{50}$ kpc) around the central galaxy, 
with isophotal magnitudes in the range $23.3 \leq B_j \leq 24.6$
($ -21.2 \leq M_B \leq -19.8$ at $z=0.75$), and two additional ones of the 
same luminosity within a distance of $60''$. All these bright galaxies have similar
SEDs and belong to the blue population mentioned above. Compared to 
low and medium-redshift rich clusters, such as A2218 or A370, the number of bright 
galaxies ($M_B \leq -20 $) found within a radius of about $300 h^{-1}_{50}$ kpc 
around the central galaxy is lower: only 6 galaxies compared to 24 in A2218 and
22 in A370, under the same conditions. But A2218 and A370 are among 
the richest Abell clusters. The difference reduces when it is compared 
to high-redshift clusters from the
EMSS Survey (Gioia \& Luppino, 1994), such as MS1054.4-0321 and MS1137.5+6625,
with tentative redshifts $z=0.81$ and $z=0.65$ respectively, where the expected
number of bright galaxies is 13 to 14 (assuming a contamination of 
about 30\%). Again, these clusters are very rich and strong X-ray emitters. 
We conclude that the cluster detected is probably not very rich. The main difference 
compared to A2218 or A370 is the evolutionary state of the bulk population of 
galaxies: even the brightest members seem to be undergoing an episode of active
star-formation, or have experienced a burst just a few Gyr ago.

\begin{figure}
\centerline{\psfig{figure=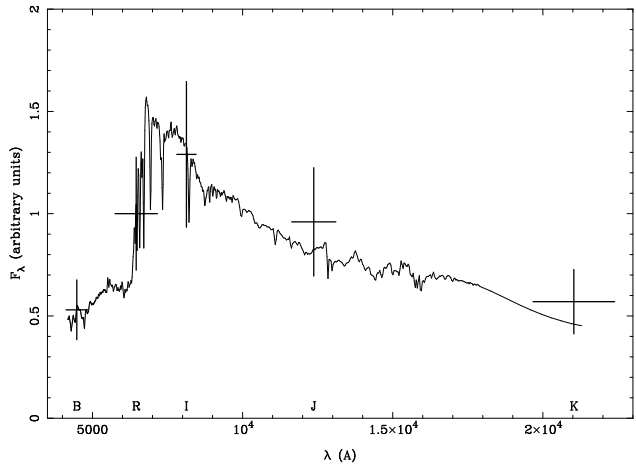,height=7.5cm}}
\caption [] { SED of the central galaxy of the distant cluster, 
with $F_{\lambda}$ normalized at 1 in R, superimposed to the best fit 
synthetic spectrum of a star-formation burst 1.3 Gyr old. The redshift
of the synthetic spectrum corresponds to the best fit for this SED, 
$z = 0.7$}
\end{figure}

\section{Photometric study of other redshift systems at z = 0.28, 
z $\simeq$ 1.2 and z $\sim$ 1.8}
There are three background galaxies in the field with
spectroscopic redshifts around 0.28 (Bonnet et al. 1993), so we can wonder 
if a concentration exists at such a redshift. These galaxies are
identified as G1, G2 and G3 in figures 1 and 2. In all the cases, the
SEDs derived from photometry are in good agreement with the observed 
spectra within the common wavelength domain. The stellar population 
in G1 corresponds to a 9 Gyr old burst of star-formation, according to the 
minimization procedure to find the best fit by a synthetic spectrum. On the 
contrary, the SEDs of G2 and G3 are dominated by a younger population, such
as a burst of star-formation about 3 Gyr old in both cases. Although these
three galaxies are among the brightest objects in the field, there is no
photometric evidence for a cluster at such a redshift. Only 26 objects are
compatible with this redshift, including the arclet-candidate A2 (see Sect. 
8), and their 2D distribution does not show any significant
concentration in the projected number density. Besides, a cluster at $z=0.28$
would have to appear more clearly in the images than a cluster at $z=0.75$. 
All these results 
point to the existence of only a small group of which G1, G2 and G3 are 
the brightest members. In addition, according to our data, the population of galaxies at 
$z=0.75 \pm 0.05$ does not show any particular alignement which could be 
interpreted as an induced effect of gravitational shear by a foreground lens. 
This result is also in good agreement with the results of Mellier et al. (1994). 


   The total number of objects compatible with $z=1.2 \pm 0.1$ is 62, but
22 of them are also compatible with $z=0.75$: these are faint objects for
which the photometric redshifts are poorly determined. Most of their magnitudes are
beyond the completeness limit in at least one filter. The mean projected 
number-density of such objects over the whole field is
15.4 galaxies/arcmin$^2$, when we take all the sample,
without any selection in magnitude. Three clumps appear in the projected 
number-density. The most important local maximum lies 
$12\arcsec$ eastwards and $42\arcsec$ northwards with
respect to the center of shear; it is 3.4 times the mean value, with a
contrast of about 9.5. Nevertheless, it is produced by the presence of several
faint galaxies beyond the completeness limit in all the filters, and the
contrast reduces to zero when we consider a circular region of more 
than $20\arcsec$ radius around it. When the sample is limited to objects 
brighter than J=25 (a value close to the completeness limit), only 41 objects
remain and the previous maximum disappears (see fig. 9a). Objects in this sample 
are located preferentially in two clumps, where the number-density never 
exceeds 2.5 times the mean value, so the distribution within these regions is 
quite smooth. One of the regions overlaps the center of the cluster at
 $z \sim 0.75$. Again, as in the $z \sim 0.75$ plane, the field 2
is not a blank field at $z \sim 1.2$, and the excess was already visible in
figure 5a. It is worth noting that objects belonging to this excess 
at $z \sim 1.2$ are expected to be lensed and slightly amplified by the 
cluster at $z \sim 0.75$, so 
their spatial distribution is probably the result of a magnification bias. 
Unfortunately, we cannot test on the existence of a systematic effect 
of shear on such objects because their angular size is too small. The images 
used in this work are deep enough to allow a good detection level and photometry, 
but too poor for an accurate determination of the shape parameters.

\begin{figure*}
\centerline{\psfig{figure=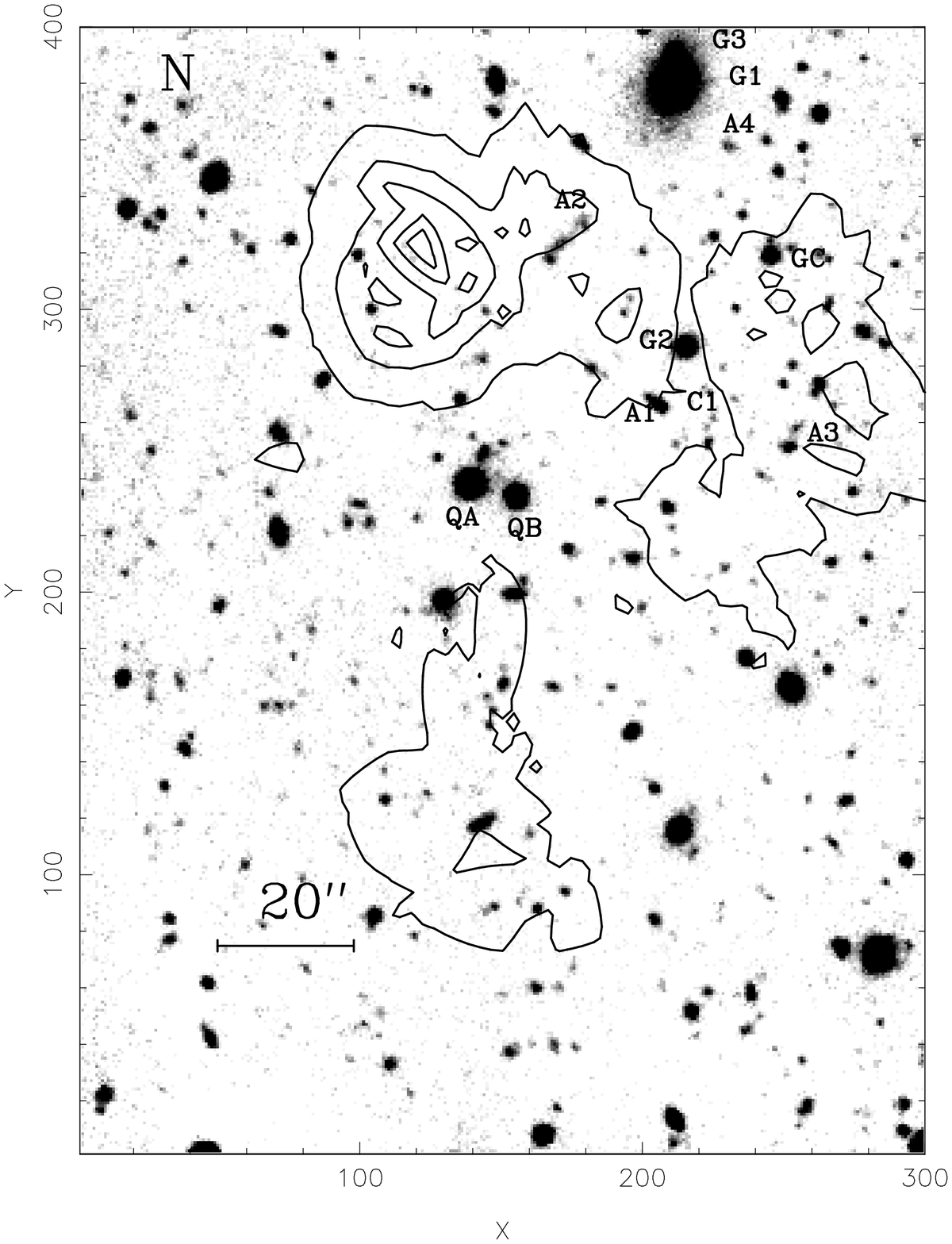,height=11.6cm}
\psfig{figure=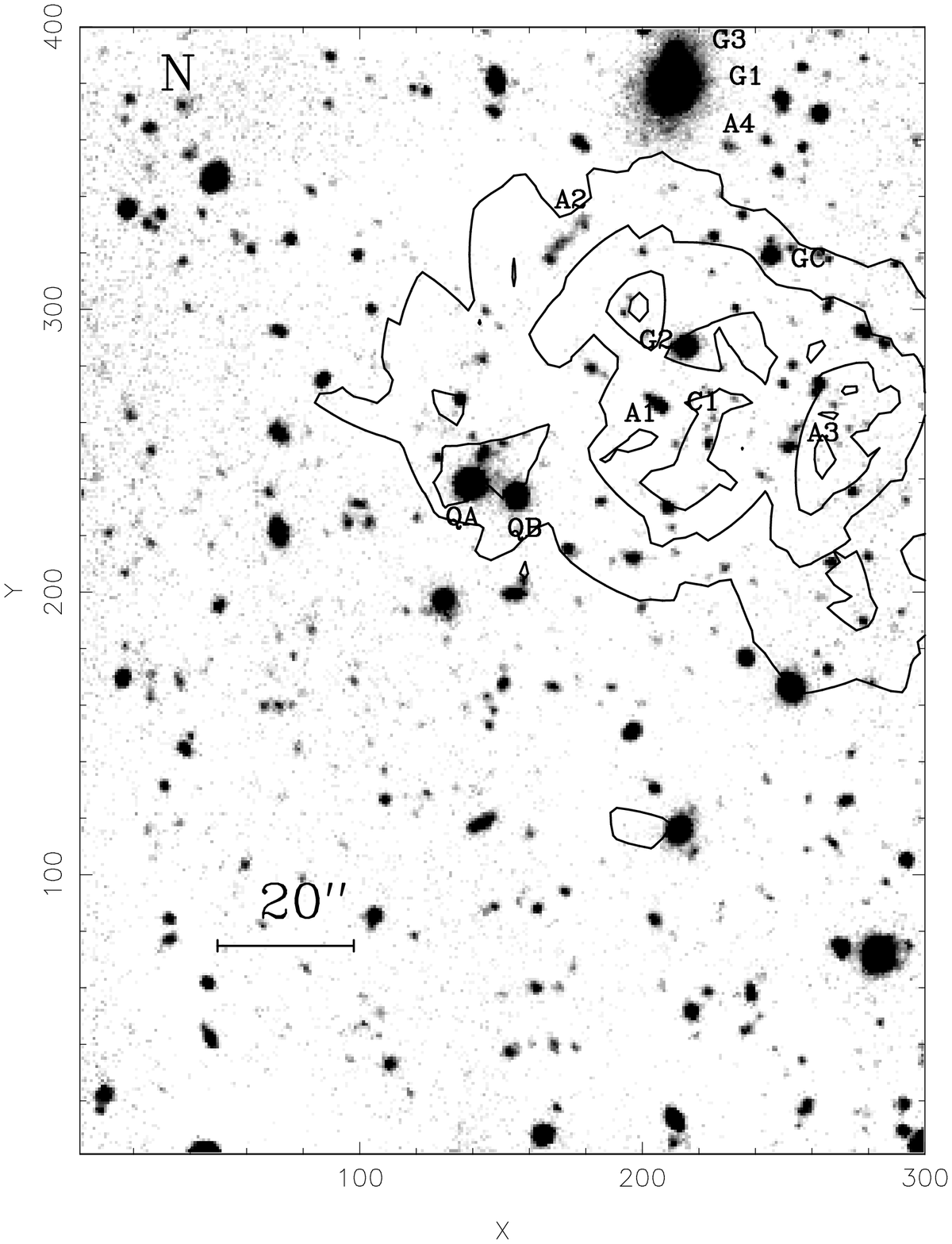,height=11.6cm}}
\caption [] { a) Isocontour plot of the projected number density of galaxies
compatible with the excess at $z \simeq 1.2$, up to the completenss magnitude
in B$_J$, superimposed on a composite image of the common field. Same comments as in
Fig. 6. b) Isocontour plot of the projected 
number density of galaxies compatible with $1.5 \leq z \leq 2.0$. Successive 
contours correspond to 0.5 times the mean value, starting at 1.
}
\end{figure*}

   The last interesting concentration is that identified with
the excess at $z \sim 1.8$, which contains most of the absorption
systems found in the spectra of the two QSOs. 44 objects are found on
the whole common field with photometric redshifts in the
interval $1.5 \leq z \leq 2.0$, and 42 of them are located within the
field 1 (see fig. 9b), within a wide region centered around the center
of shear and the main galaxy at $z \sim 0.75$. Almost all of them are fainter
than the completeness magnitudes in all the filters, so any detailed discussion 
about their 2D distribution has to be taken with caution. The maximum
in the projected number-density is 2.8 times the mean value, and the
maximum local contrast is 6, but the
2D distribution is quite smooth. Again, these objects are affected by 
the presence of the
foreground cluster, and a systematic effect of shear and magnification 
is expected for them. 

\section{The arc(let)s revisited}
Three arclet candidates were reported in the paper by Bonnet et al.
(1993) and another one was added by Mellier et al. (1994). Figure 10 shows 
the morphology of these arclets in the different
filters. They are all located close to the excess of galaxies at $z \sim
0.75$. A2 and A3 show a complex morphology, whereas A1 and A4 are
quite compact. According to the SEDs deduced from photometry,
A1 and A4 seem to be cluster members. A1 is the fourth brightest 
cluster galaxy identified in this field and its SED is shown in 
figure 11a. The automatic procedure to fit the best synthetic spectrum to 
A1 and A4 has found that both SEDs are compatible with a recent burst
of star formation, such as an E-type galaxy 1 and 2 Gyr old, 
respectively, at a redshift of $\sim 0.75$. The SED of the arclet A3 is
rather similar (an E-type galaxy 2.8 Gyr old), but the best fit for the
redshift is lower, $z \sim 0.6$, according to the automatic procedure
and marginally compatible with the cluster redshift.
Figures 11b and 11c show the best automatic fit for arclets A3 and A4,
respectively. So, these three arclets are more likely elongated cluster 
or foreground galaxies rather than gravitational arclets.

   The arclet A2 is more complex and elongated. Three different regions 
can be distinguished 
in it: two blue regions, named $A2_1$ and $A2_3$ in figure 11d and Table 3, 
and a central region ($A2_2$) which is the reddest. The colors of regions 
$A2_1$ and $A2_3$ are quite similar, whereas $A2_2$ has more flux in the K' band.
Two different solutions for the redshift seem possible. The most reliable is 
that of a low redshift object (with $z \leq 0.55$), and the automatic fit of the
best spectrum in the range $0 \leq z \leq 0.6$ gives a solution at 
$z=0.31$ for the three regions. So, it could be a member of the group at 
$z \sim 0.3$. The best synthetic spectra at such a redshift 
correspond to a recent burst of star formation ($A2_1$ and $A2_3$) or an old 
late-type S galaxy ($A2_2$). The relevant spectral feature is the presence of a
break between the filters B$_J$ and R in all the three regions, which corresponds
to the $4000 \AA$ break in the hypothesis of $z \sim 0.3$. In $A2_2$, the
flux increases slightly between J and K'. If this is not due to a chance alignement
along the line of sight (i.e. with a M star), another possible different solution 
is that of a high redshift object with the $4000 \AA$ break between J and K'. 
The main break in the spectrum would correspond to the
Lyman break in this case, and the best fit for the blue and the red regions 
would be $z \sim 3.6-3.8$. The difference between blue and red regions in the 
high redshift hypothesis is due to a difference in the age of the stellar 
populations. It is worth noting that the high redshift hypothesis can be 
tested spectroscopically, because $Ly\alpha$ is expected to lie at
$\lambda \sim 5600-5900$ \AA.

\begin{table}
\caption[]{Photometry of arclets.}
\begin{flushleft}
\begin{tabular}{lrrrrrl}
\hline\noalign{\smallskip}
Id. & $\mu_B$ & B$_J$-R &R-I &I-J &J-K' & Photometric \\
& max & & & & & redshift\\
\noalign{\smallskip}
\hline\noalign{\smallskip}
$A1$ &24.79&0.56&1.17&0.50&1.15&0.60-0.80 \\
$A2_1$ &25.82&1.43&0.55&0.51&1.74&0.0-0.55/$\sim$3.6-3.8 \\
$A2_2$ &26.11&1.39&0.65&0.51&2.18&0.18-0.55/$\sim$3.6-3.8\\
$A2_3$ &26.20&1.30&0.77&0.30&1.66&0.0-0.55/$\sim$3.6-3.8 \\
$A3$ &26.14&1.64&0.57&0.72&1.71&0.07-0.65 \\
$A4$&26.20&0.87&1.22&0.53&1.67&0.58-0.90 \\
\noalign{\smallskip}
\hline
\end{tabular}
\end{flushleft}
\end{table}

\begin{figure}
\centerline{\psfig{figure=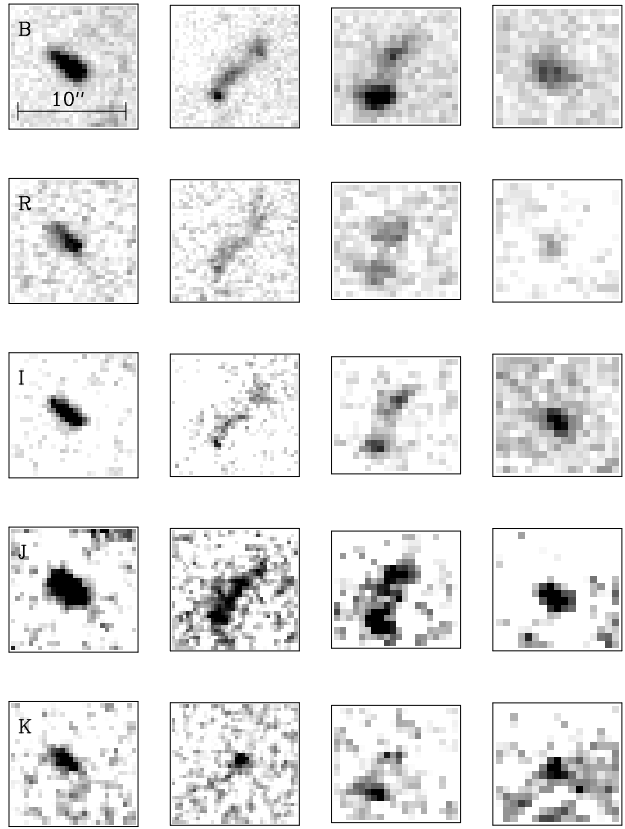,height=11.6cm}}
\caption [] {Zoom of the field in the close neighbourhood of
the four arclet candidates, A1 to A4 from right to left,
as seen in the different filters.
}
\end{figure}

\begin{figure*}
\centerline{\vbox{
\psfig{figure=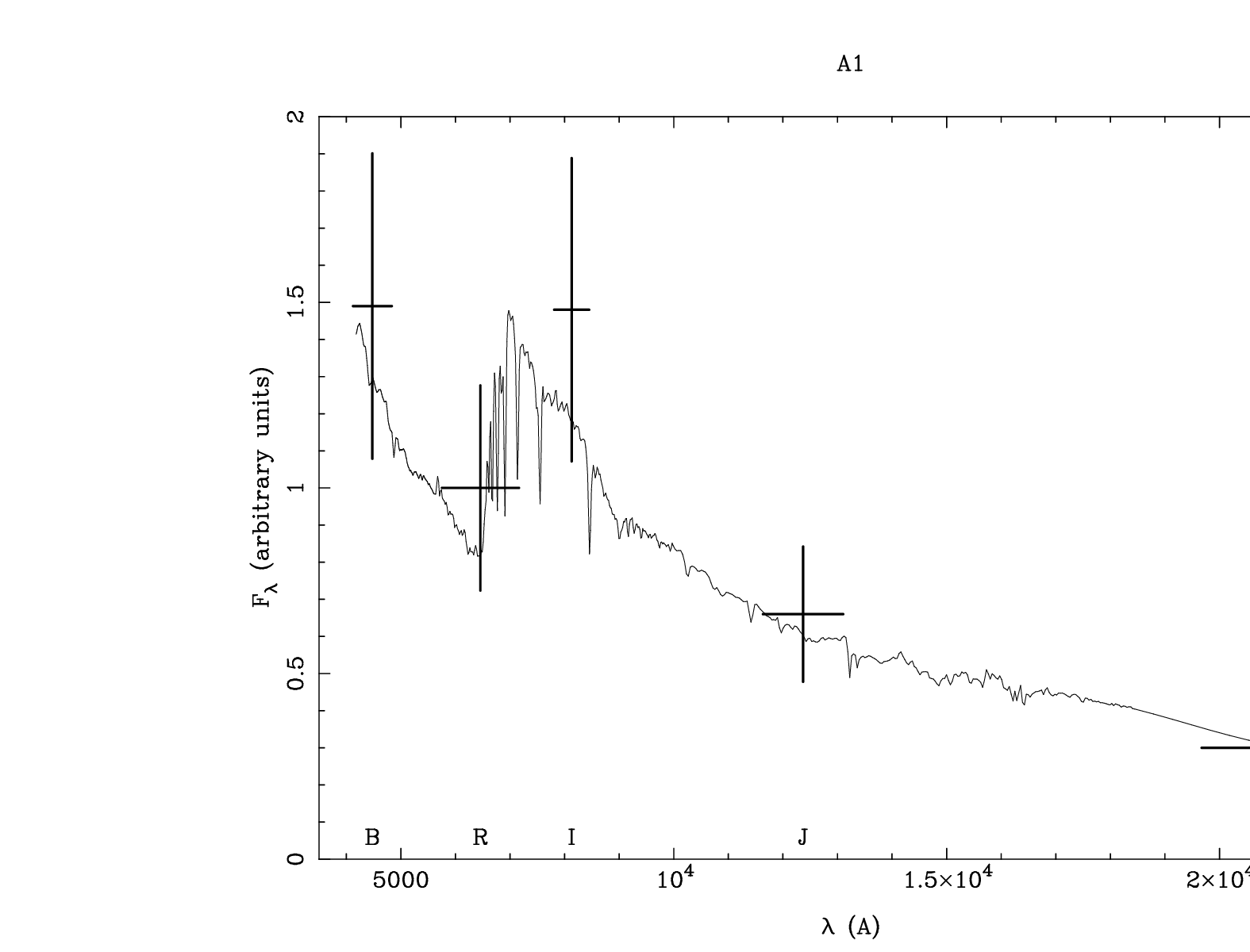,height=5.5cm,width=8.9cm}
\psfig{figure=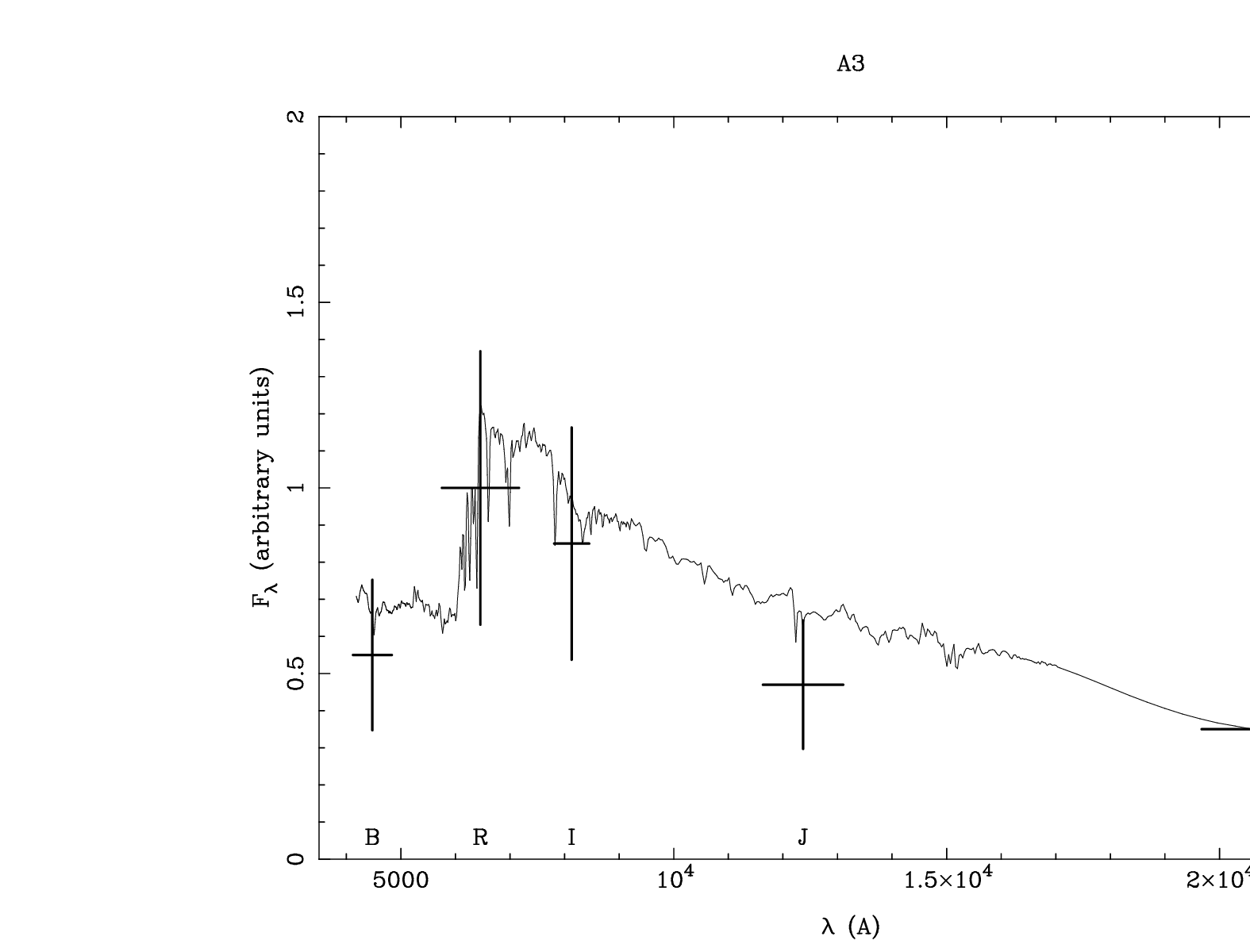,height=5.5cm,width=8.9cm}
\psfig{figure=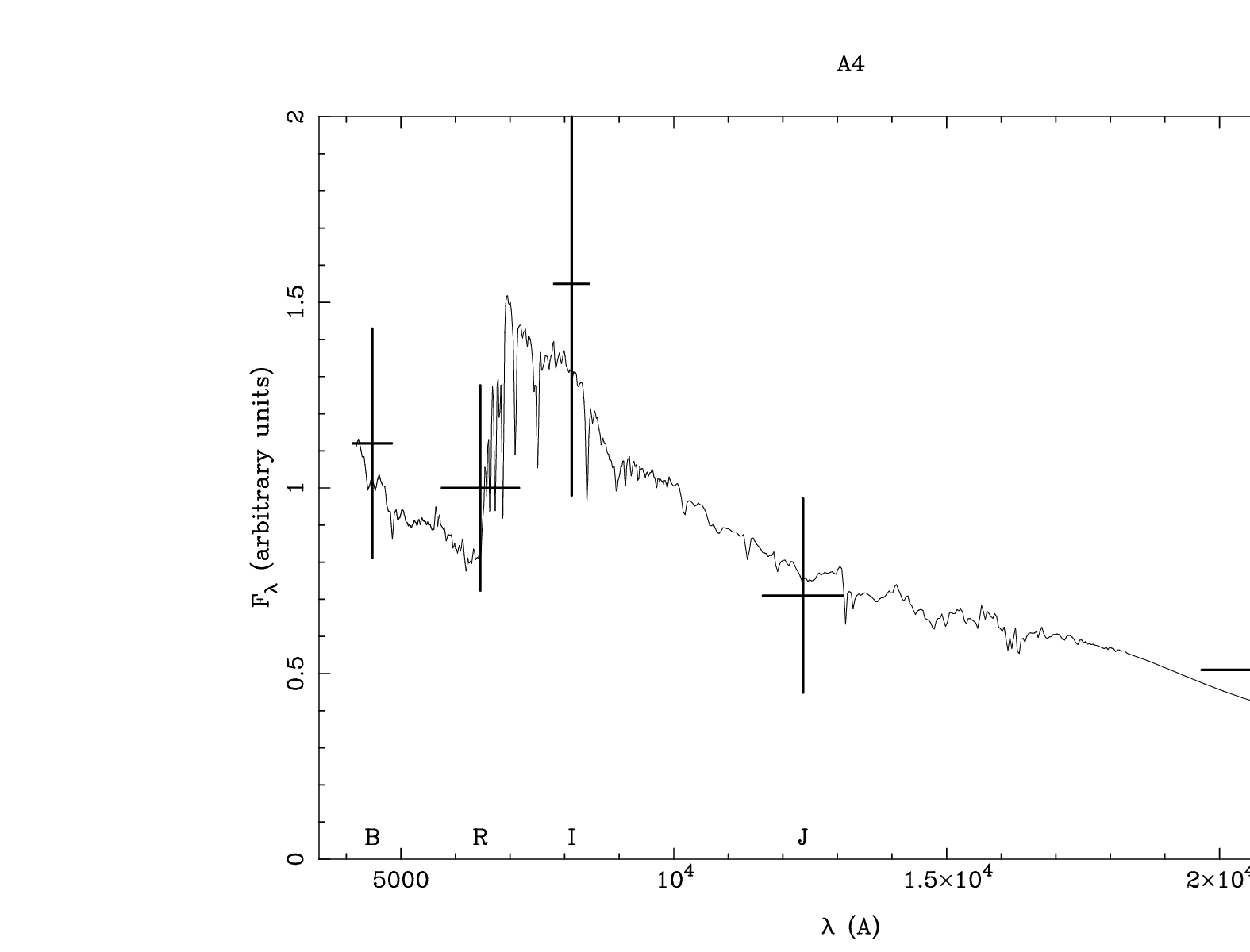,height=5.5cm,width=8.9cm}}
\psfig{figure=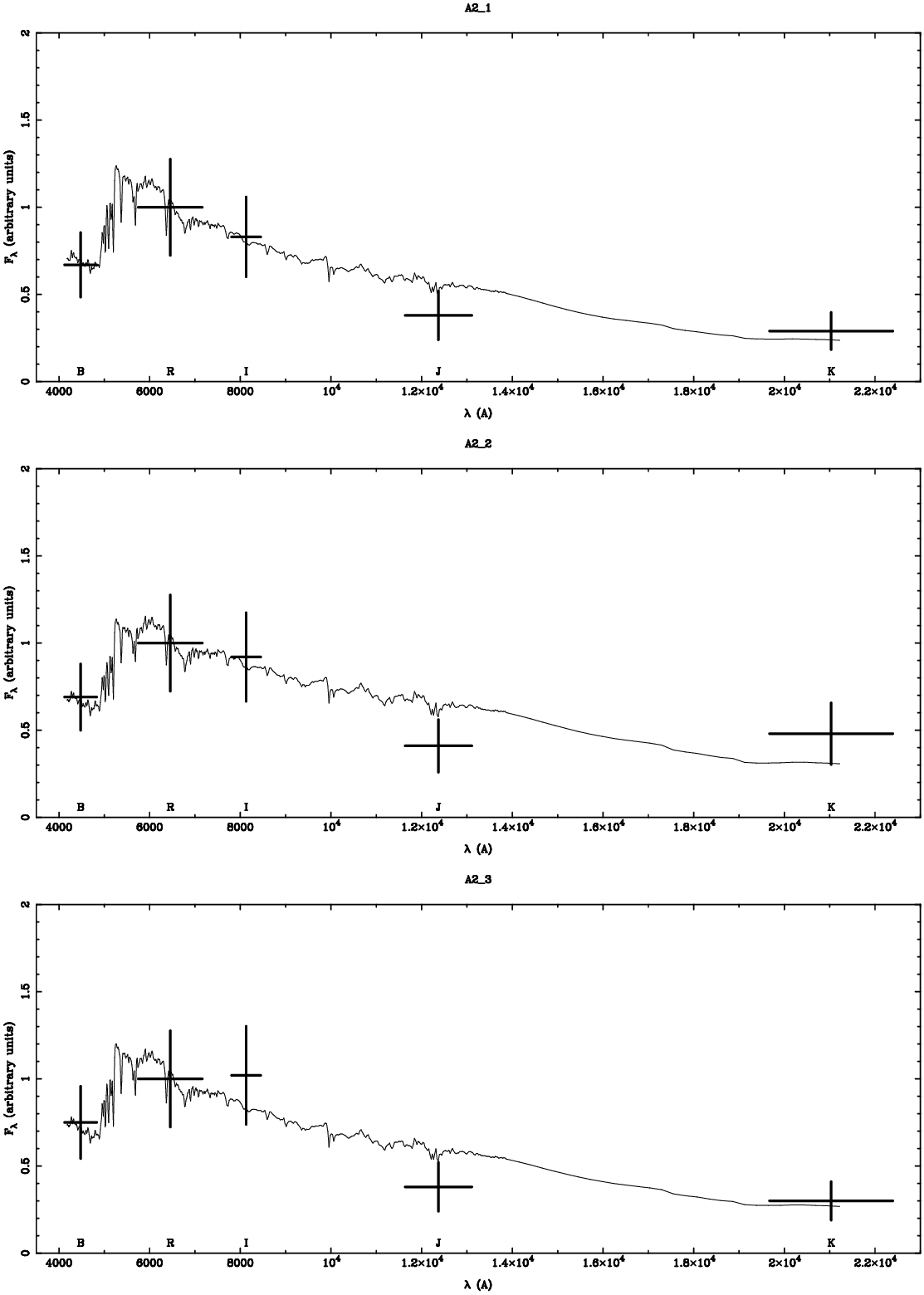,height=12.5cm,width=11.0cm}}
\caption [] {SEDs derived from photometry for the four arclets, superimposed to
the best-fit synthetic spectra: a) A1, with the spectrum of an E-type galaxy 1.0 Gyr
old at $z=0.75$; b) A3, with the spectrum of an E-type galaxy 2.8 Gyr old at 
$z=0.61$; c) A4, with the spectrum of an E-type galaxy 2.1 Gyr old at $z=0.75$;
d) $A2_1$, $A2_2$, and $A2_3$, with the best-fit spectra at $z=0.31$ (see text
for details). }
\end{figure*}

\section{Discussion and conclusions}
The present work suggests that Q2345+007 is probably
the result of a very complex gravitational lens involving
several mass concentrations at different redshifts. We confirm the presence 
of a group of galaxies 
rather than a cluster at $z=0.28$. The main result is the identification and
characterization of a distant medium-to-rich cluster at $z=0.75 \pm 0.08$,
which appears as an excess in the redshift distribution with
respect to a blank field. The center of the cluster is located
close to the center of the shear-field. Bonnet et al. (1993) 
already analysed this field and they obtained a good fit of the
shear pattern by a singular isothermal sphere (SIS) with 
$\sigma = 1200 km s^{-1}$ at z=1. Assuming that the bulk of the 
mass in the lens is associated with the cluster, the velocity dispersion 
required in order to produce the same mean shear pattern should range from 
$620$ to $905 km s^{-1}$, the more probable value 
beeing $790km s^{-1}$. This value was estimated 
with the following assumptions: the mass-distribution follows a 
SIS, the redshift of the lens is $0.65 \leq z \leq 0.85$, and the 
shear field measured comes mainly from sources around $z \simeq 1.2$ 
and $z \sim 1.8$. The 
resulting K-corrected M/L ratios in the inner $300 h_{50}^{-1} kpc$ radius
are $M/L_B \simeq 250 M_{\odot}/L_{\odot}$ and 
$M/L_J \simeq 40 M_{\odot}/L_{\odot}$. 

   Most cluster-member candidates at $z \sim 0.75$ show SEDs with 
signatures of a recent episode of star-formation. About 50\% of them 
are undergoing a star-forming process whereas the others are more 
likely burst-systems where the 
star-formation stopped between 1 and 3 Gyr ago. An old population of stars
is present in about 50\% of galaxies. These secondary-burst systems are 
similar to the population of galaxies that Barger et al. (1995)
find in several medium-redshift clusters (about 30\% of the population
in such clusters) but, in our case, more than 40 \% of the population
within the completeness limit in B$_J$ shows bursts, including the 10
brightest galaxies. Only 2 galaxies 
from this sample show SEDs corresponding to a pure old-population 
of stars, such as a non-evolved E-type galaxy, and they are relatively
faint. These results indicate that the evolutionary state of the 
bulk population of galaxies is different from what is observed
for medium-rich clusters at lower redshifts. The existence of short
bursts of star formation at redshifts $0.5 \leq z \leq 1$ involving 
low-mass galaxies has been suggested by different authors to explain
the results on deep galaxy counts (for example, Cowie et al. 1991, 
Lilly 1993, Babul \& Ferguson 1995).
According to our results, these star-formation processes could 
also affect the bulk population of galaxies in clusters within this 
redshift range.

   Two other excesses of galaxies appear in the field of the double
QSO with respect to the blank field, which points out the existence of
clustering at redshift-strips $z \simeq 1.2$ and $z \sim 1.8$. In both 
cases, the 2D distribution in projected number-density is quite smooth,
and the objects tend to be located around or close to the cluster 
at $z \sim 0.75$. These objects are faint, beyond the completeness 
magnitude in at least one filter in most cases, especially for the 
excess at $z \sim 1.8$. But, 
even if the 2D distribution has to be taken with caution, 95\% of 
objects compatible with  $z \sim 1.8$ are located in the field 1.
As they suffer the gravitational amplification induced by the foreground
cluster, they are probably detected in the field 1 just because the
magnification bias tend to enhance their total luminosities
in this region compared to the blank field. A systematic
effect of shear should appear on this population if this identification is
correct. A more detailed study is needed to measure the shear on this
pre-selected sample of high-redshift objects, using deep and well sampled 
images. In any case, we can already say that the typical magnitudes of 
objects belonging to the
redshift-strips at $z \sim 1.2$ and $z \sim 1.8$ (B$_J$  $\geq 27$) make 
them compatible with the population used by Bonnet et al. (1993) 
to determine the shear field. The detection of a so faint shear-effect 
was probably made easier (or even possible) by the existence of clustering 
along the line-of-sight. Concerning the double QSO, the existence of several 
excesses at different redshift-strips indicates that
we are probably dealing with a complex lens. The high separation
of the 2 images is the combined effect of these lens-planes, including the 
faint galaxy in the close neighbourhood of the fainter QSO reported by 
Fischer et al. (1994). As a final remark, we can mention that none of the
relatively blue objects detected $10\arcsec$ around the double QSO has
a photometric redshift assigned, because their SEDs are quite different
from the models adopted in this paper. Concerning the complexity
of the lens, a similar result was found by Angonin-Willaime et al (1994) 
in the field of the double QSO 0957+561,
where the main lens is the sum of a bright gE galaxy and of the cluster of
galaxies associated with it. The field of QSO 0957+561 is also contaminated 
by a background
group of galaxies identified spectroscopically, and many rather bright
foreground galaxies. Their effects on the high separation of the 2 images 
are not clear and probably not dominant, but it should affect strongly 
the shear field of the faint background galaxies.

Among the arclets identified by Bonnet et al. (1993) and by
Mellier et al. (1994), A1 and A3 are more likely elongated cluster members
rather than gravitational arclets, whereas A4 seems to be a foreground
object, only marginally compatible with the cluster redshift. 
No other elongated arclet-candidates are visible. The only
exception is A2, for which a high redshift solution exists, at
$z \sim 3.6-3.8$, although a more reliable one identifies it as a 
member of the group at $z \sim 0.3$. It is important for the lens
modeling to confirm the high redshift hypothesis for this object,
and it can be done because $Ly\alpha$ is expected at $\lambda 
\sim 5800 \AA$.

   One of the priorities for the near future is to complete an 
extended spectroscopic survey on this field to confirm the predicted 
distributions in redshift. Recently, some additional spectroscopic data have 
been obtained at the CFHT, and the reduction is going on. The presence of
strong emission-lines is expected in the visible range for those brightest
cluster-member candidates which are undergoing an active process of star
formation, making possible the measure of the redshift. It is worth
noting that the study of the SEDs from B$_J$ to K' allows
not only an estimate of the photometric redshift but also a
determination of the optimal wavelength and feasibility for a successful 
measure of the spectroscopic redshift. In a more general context, 
we expect to check the reliability of the statistical method at high-redshift. 
Combining this technique of multicolor analysis with the 
systematic studies of weak-shear fields, and lensed QSOs in particular,
is a very promising way to characterize the visible counterparts
of large scale inhomogeneities.
 
\acknowledgements We thank Y. Mellier, for a careful reading
of the manuscript, and H. Bonnet, B. Fort, G. Mathez
and L. Van Waerbeke for their helpful comments and discussions all 
along this work. We are specially indebted to M. Dantel-Fort for 
her assistance during data reduction and to J. Bezecourt for his 
help in getting the filter transmissions.
Part of this work was supported by the French Centre National de
la Recherche Scientifique, by the French Groupe de Recherche en
Cosmologie and the EC HCM Network CHRX-CT92-0044.
JMM is grateful for financial support from an
A.D.I. Research Grant and from the
Ministeri d'Educaci\'o del Govern d'Andorra. GB acknowledges
the CE for partial support through the ALAMED contract no 
Cil-CT93-0328VE.

\appendix
\section{Appendix. Photometry of the Double QSO}
Table 4 gives the magnitudes and colors of the 2 components of
the double quasar. The
photometric results in the optical region are compatible with those
published by Fischer et al. (1994) within the errors, and no other new
candidate to lens galaxy appears in J and K'. Fischer et al. (1994) found a 
magnitude difference between the two components of $1.31 \pm 0.05$ 
in B$_J$ and $1.18 \pm 0.05$ in R, from the composite 
images taken between 1984 and 1993. The difference in magnitude
is more important in the near-IR but this could be explained by 
the variability of the quasar between the optical and the near-IR runs. 
The colors are also very similar within the errors, except for I-J.

\begin{table}
\caption[]{Photometry of the double QSO }
\begin{flushleft}
\begin{tabular}{lrrrrrr}
\hline\noalign{\smallskip}
& B$_J$ & R & I & J & K' \\
\noalign{\smallskip}
\hline\noalign{\smallskip}
A & 19.47 & 18.94 & 18.48 & 18.01 & 16.96 \\
B & 20.71 & 20.13 & 19.62 & 19.56 & 18.47 \\
\noalign{\smallskip}
\hline\noalign{\smallskip}
${\Delta}m$ & 1.24 & 1.19 & 1.14 & 1.55 & 1.51 \\
\noalign{\smallskip}
\hline
\end{tabular}
\end{flushleft}
\end{table}

\end{document}